\documentclass[lettersize,journal]{IEEEtran}
\usepackage{amsmath,amsfonts}
\usepackage[noend]{algpseudocode}
\usepackage{array}
\usepackage[caption=false,font=normalsize,labelfont=sf,textfont=sf]{subfig}
\usepackage{textcomp}
\usepackage{stfloats}
\usepackage{url}
\usepackage{verbatim}
\usepackage{graphicx}
\def\BibTeX{{\rm B\kern-.05em{\sc i\kern-.025em b}\kern-.08em
    T\kern-.1667em\lower.7ex\hbox{E}\kern-.125emX}}
\usepackage{balance}

\usepackage[style=ieee, doi = false, isbn = false, url = false]{biblatex} 
\usepackage[dvipsnames]{xcolor}
\graphicspath{{Images/}}
\usepackage{import}
\usepackage{algorithm}

\algrenewcommand\algorithmicindent{0.65em}%
\algnewcommand{\IIf}[1]{\State\algorithmicif\ #1\ \algorithmicthen}
\algnewcommand{\ElseIIf}[1]{\algorithmicelse\ #1} 
\algnewcommand{\EndIIf}{\unskip\ \algorithmicend\ \algorithmicif}
\algrenewcommand{\alglinenumber}[1]{\footnotesize#1:}
\usepackage{setspace}
\usepackage{amssymb}
\usepackage{threeparttable}
\usepackage{booktabs}
\usepackage{tabularx}
\usepackage{mathtools}
\usepackage{enumitem}
\usepackage{relsize}

\begin{document}
\title{A Synchrophasor Estimator Characterized by Attenuated Self-Interference and Robustness Against DC Offsets: the DCSOGI Interpolated DFT}
\author{César García-Veloso\thanks{Manuscript created January, 2025; The project that gave rise to these results received the support of a fellowship from ”la Caixa” Foundation (ID 100010434). The fellowship code is LCF/BQ/DR20/11790026. The project also received the support of the Schweizerischer Nationalfonds (SNF, Swiss National Science Foundation) via the National Research Programme NRP 70 “Energy Turnaround” (project nr. 197060).}\thanks{César García-Veloso is with the Department of Electrical Engineering, University of Seville, 41092 Seville, Spain and also with the Distributed Electrical Systems Laboratory, École Polytechnique Fédérale de Lausanne, 1015 Lausanne, Switzerland (e-mail: cgveloso@us.es).},~\IEEEmembership{Graduate Student Member, IEEE}, Mario Paolone\thanks{Mario Paolone is with the Distributed Electrical Systems Laboratory, École Polytechnique Fédérale de Lausanne, 1015 Lausanne, Switzerland (e-mail: mario.paolone@epfl.ch).},~\IEEEmembership{Fellow, IEEE}, José María Maza-Ortega\thanks{ José María Maza-Ortega is with the Department of Electrical Engineering, University of Seville, 41092 Seville, Spain (e-mail: jmmaza@us.es)},~\IEEEmembership{Member, IEEE}}

\maketitle

\begin{abstract} The second-order generalized integrator (SOGI), which can be used to attenuate the self-interference of the fundamental tone, is unable to reject DC offsets on the input signal. Consequently, the performance of any SOGI based synchrophasor estimation (SE) technique might be compromised in the presence of such DC components. The current work presents a SE algorithm which adopts and enhanced SOGI formulation, robust against DC, combined with a three-point IpDFT and a Hanning window. Two alternative formulations relying respectively on the use of two and three nominal fundamental period observation windows are proposed and assessed for simultaneous compliance with both phasor measurement unit (PMU) P and M performance classes. This is done by means of a simulated environment where all the operating conditions defined by the IEC/IEEE Std. 60255-118-1-2018 are evaluated simultaneously combined with a 10\% static DC and under two different noise levels. Furthermore, both formulations adopt a dedicated mechanism for the detection and correction of low amplitude 2\textsuperscript{nd} harmonic tones to ensure their compliance with the standard can be maintained in the presence of such disturbances even under off-nominal frequency conditions. Finally the resilience of both methods against multiple simultaneous harmonic interferences is also analyzed.
\end{abstract}

\begin{IEEEkeywords}
IEC/IEEE Std 60255-118-1-2018, second-order generalized integrator, DC offset rejection, discrete Fourier transform, interpolated DFT, phasor measurement unit, synchrophasor estimation.
\end{IEEEkeywords}

\section{Introduction}\label{doc_sec_1}
All M-class PMUs must meet the requirements set in \cite{PMU_Measurement_60255-118-1-2018} for out-of-band interferences (OOBI). This represents an extremely demanding condition for DFT-based synchrophashor estimation (SE) techniques when short observation windows are adopted. This is because the latter results in an increased spectral proximity and interference among the different components of the signal. While DFT-based PMUs represent the majority of commercial devices \cite{Kamwa2014_WFRA_PMU_Algos}, very few DFT SE algorithms seek to comply with both P and M performance classes simultaneously, and among those that do, typically at least three nominal cycle windows are adopted \cite{Derviskadic2018i-IpDFTforSynchrophasorEstimation, Derviskadic2020_iIpDFT_FPGA,Frigo2019_HTIpDFT,Song2022_FiIpDFT,Song2022_FiIpDFT_Addendum,Veloso2023_SOGI_IpDFT_PT,Veloso2023_TIM}. Their main difference lies in the way the self-interference of the fundamental tone is handled i.e. how the mutual interaction between its positive and negative images is managed. Indeed, those applications where a fast response is required, such as protections \cite{Meier2017_uSPhasors_Apps}, would benefit from the reduced latency offered by the adoption of shorter windows. However, the coarser frequency resolution and increased spectral interferences between signal components they cause also heavily complicate the estimation.

To the best of our knowledge, only \cite{Frigo2019_HTIpDFT} and the recent work in \cite{Shan2023_eIpD2FT-SE_for_M_class} have been shown to comply with the OOBI test using a two-cycle window. 
In \cite{Frigo2019_HTIpDFT} the use of two parallel infinite impulse response (IIR) Hilbert filter banks of different orders is proposed to derive an approximate one-sided spectrum analytic signal. The selection between the higher (more accurate) and lower order (faster) filter bank is done via a heuristically tuned transient detection mechanism based on a 3\textsuperscript{rd} order Butterworth low-pass filter. The method, named HT-IpDFT is shown to comply with the OOBI test with a two nominal cycle observation window. However, it presents limitations:
\begin{itemize}
    \item Even though the provided results showcase compliance with the OOBI test, the authors indicate that 'the efficiency of the interference compensation routine cannot be totally guaranteed' \cite[p.~3476]{Frigo2019_HTIpDFT} under the short window.
    \item The method is non-compliant with the 1\% harmonic distortion (HD) P-class test while using a two nominal window as no detection mechanism for such low interferences is used.
    \item Resilience against DC offsets is not investigated nor are the filters DC gains explicitly discussed.    
\end{itemize}
Regarding \cite{Shan2023_eIpD2FT-SE_for_M_class}, a computationally demanding method is adopted based on a dynamic phasor model relying on a large data buffer to determine the number of narrowband components within the signal. It employs a three-step process involving: (i) the use of an algorithm based on random matrix theory to determine the number of narrowband components, (ii) the estimation of their frequencies through ESPRIT \cite{Roy1989_ESPRIT}, and (iii) their characterization through the Interpolated Dynamic DFT (IpD2FT) algorithm \cite{Petri2014_IpD2FT}.

In this paper, which is an extension of \cite{Veloso2024_SOGI_IpDFT_SGSMA}, a different approach is proposed. In \cite{Veloso2024_SOGI_IpDFT_SGSMA} an alternative formulation of the SOGI-IpDFT, described in \cite{Veloso2023_SOGI_IpDFT_PT}, capable of maintaining its simultaneous compliance with the P and M performance classes \cite{PMU_Measurement_60255-118-1-2018} while reducing the observation window from three to two nominal cycles was presented. The current paper includes the following contributions and enhancements with respect to the previously published research:
\begin{itemize}
    \item Substitution of the SOGI quadrature signal generator (QSG) filter with the DCSOGI-QSG, to provide both previous SE techniques \cite{Veloso2024_SOGI_IpDFT_SGSMA,Veloso2023_SOGI_IpDFT_PT} with Static DC Immunity and allow them to correctly detect and suppress OOBI interferences even in the presence of DC offsets. As highlighted in \cite{Karimi-Ghartemani2012_DCSOGI} signals can be affected by DC offsets due to several reasons such as saturation of measurement equipment, analog-to-digital (ADC) conversion or faults.
    \item The inclusion of a dedicated 2\textsuperscript{nd} harmonic detection mechanism so both variants of the algorithm, using respectively either three or two nominal period observation windows, can identify and compensate low amplitude 2\textsuperscript{nd} harmonic tones. This also prevents the false interpretation of other phenomena as 2\textsuperscript{nd} harmonic interferences which previously affected and limited the performance of the method in \cite{Veloso2024_SOGI_IpDFT_SGSMA}.  
    \item Analysis of the performance of both formulations against multiple simultaneous harmonic interferences.
\end{itemize}
The remainder of the paper is structured as follows. Section \ref{doc_sec_2} presents the fundamentals of the new DCSOGI-QSG filter. Section \ref{doc_sec_3} describes the resulting updated SE formulation both for the two and three nominal cycle variants of the algorithm. Section \ref{doc_sec_4} details the tuning of the interference detection mechanisms used by the DCSOGI-IpDFT along with the assessment of the maximum number of iterations used for their correction. Section \ref{doc_sec_5} presents the performance evaluation based on the P and M classes defined in \cite{PMU_Measurement_60255-118-1-2018} of both variants as well as their behavior against multiple simultaneous harmonic interferences. Finally, Section \ref{doc_sec_6} concludes the paper.\vspace{-3mm}

\section{DCSOGI-QSG: DC Offset Robust SOGI-QSG}\label{doc_sec_2}
\begin{figure}[!t]
\centering
\includegraphics[width=\linewidth]{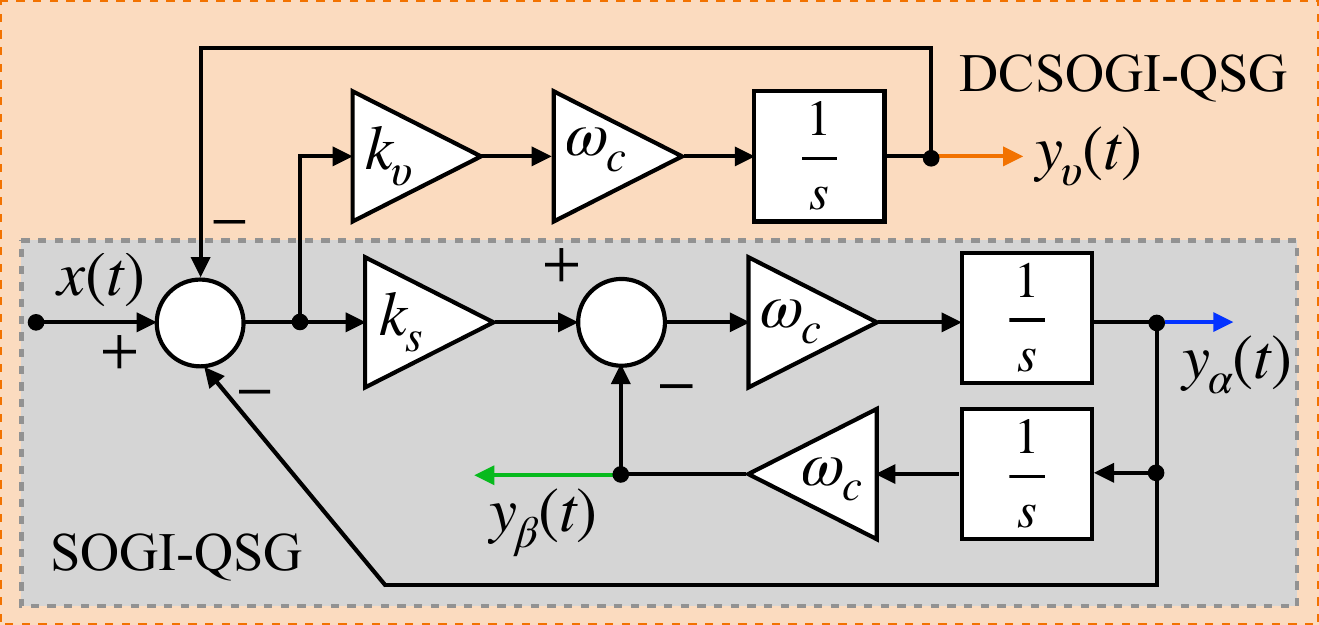}
\caption{Block Diagrams of the DCSOGI-QSG and SOGI-QSG ($k_\upsilon = 0$) for a fixed centre frequency $\omega_c = 2\pi f_c$.}
\label{fig_DCSOGI_QSG_FlowChart}
\vspace{-5mm}
\end{figure}
\begin{figure}[!t]
\centering
\includegraphics[width=\linewidth]{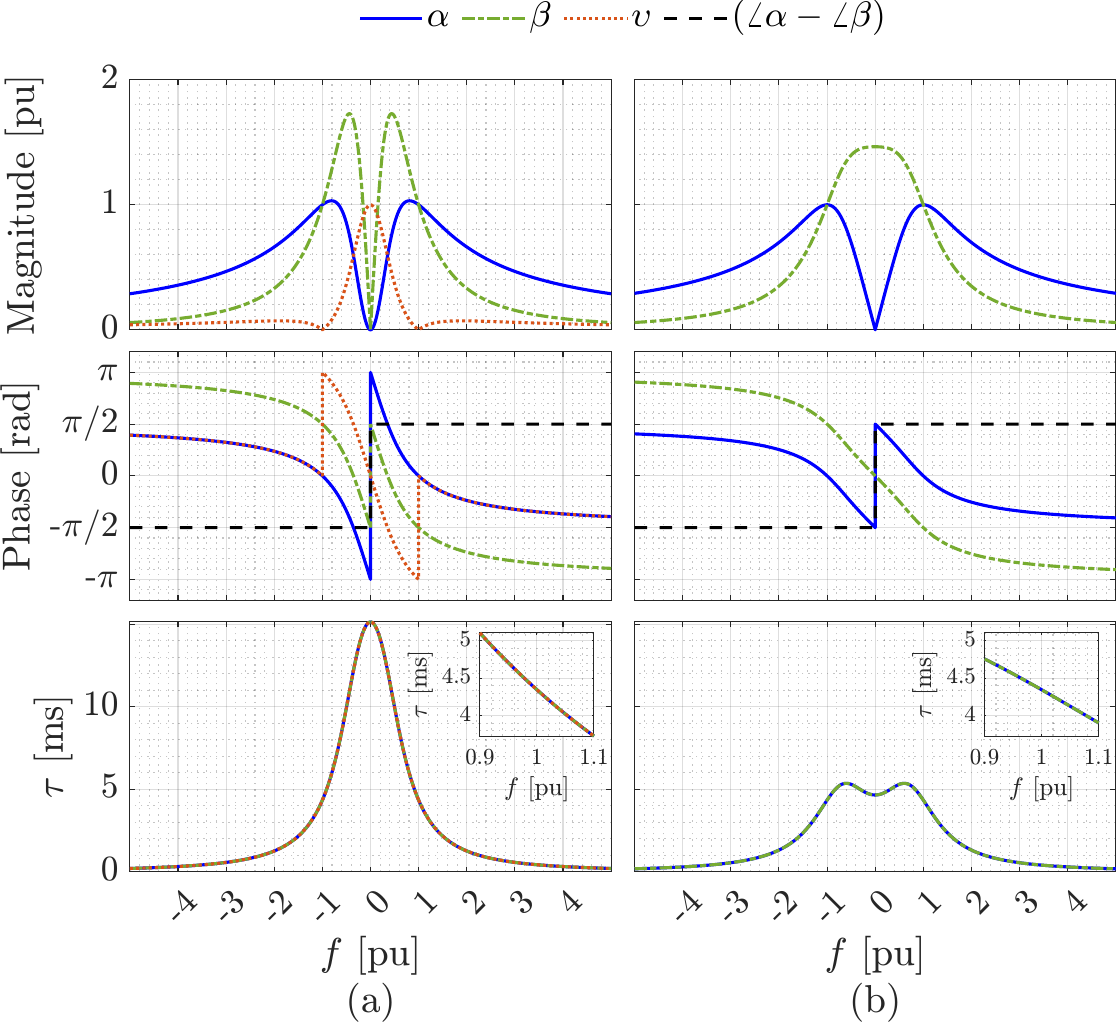}
\vspace{-5mm}
\caption{Frequency Response of the DCSOGI-QSG (a) and SOGI-QSG (b) filters for a normalized frequency $f\text{[pu]} = f/f_c$.}
\label{fig_DCSOGI_QSG_Filter_Resp}
\vspace{-5mm}
\end{figure}

As known, the SOGI-QSG (also referred to simply as 'SOGI' through this work) is a second-order adaptive filter \cite{Teodorescu2011GridConvertersforPVandWindPS-Ch4} capable of producing two in-quadrature outputs, $y_\alpha(t)$ and $y_\beta(t)$, from an input signal $x(t)$ as shown by the gray block diagram in Fig. \ref{fig_DCSOGI_QSG_FlowChart}. When operated with a fixed centre frequency $\omega_c = 2\pi f_c$ and although $y_\alpha(t)$ and $y_\beta(t)$ are in-quadrature, only for $f=f_c$, i.e. $f\text{[pu]} = 1$, no magnitude distortion is introduced. This is crucial since, as shown in \cite{Veloso2023_SOGI_IpDFT_PT}, the mitigation of the self-interference achieved by the use of the complex signal $\bar{y}(t) = y_\alpha(t) + jy_\beta(t)$ depends on the magnitude ratio between both components $\xi = A_\beta/A_\alpha$, with the positive and negative images of the signal spectrum scaled, respectively, by $1 + \xi$ and $1 - \xi$. Thus, if an initial estimate of the signal frequency is obtained and the magnitude distortion introduced by the filter is corrected, a major attenuation of the self-interference can be achieved. However, one flaw of the SOGI lies in its inability to reject potential DC offsets within the $\beta$ component (see Fig. \ref{fig_DCSOGI_QSG_Filter_Resp}(b)). While $\alpha$ can reject any DC offset present in the input signal, these are not only kept but amplified by $k_s$ in $\beta$ ($k_s$ being the SOGI gain). A review of different techniques proposed in the literate to address this issue, within the context of PLL architectures, is presented in \cite{Golestan2017_1PhPLL_Review}. Among them, the present works adopts the solution proposed in \cite{Karimi-Ghartemani2012_DCSOGI} due to its structural simplicity, where an additional loop '$\upsilon$' is added to the QSG structure to estimate and remove the DC component (see whole block diagram in Fig. \ref{fig_DCSOGI_QSG_FlowChart}). The resulting structure hereafter referred as 'DCSOGI-QSG' (or simply 'DCSOGI') produces three outputs, namely $y_\alpha(t)$,$y_\beta(t)$, and $y_\upsilon(t)$ from $x(t)$ characterized respectively, for a fixed $\omega_c$, by the following transfer functions:
\begin{subequations}\label{eq_DCSOGI_QSG_def}
\begin{align}
G_{\alpha}(s) = \frac{y_{\alpha}(s)}{x(s)} =  \frac{k_s\omega_{c}s^{2}}{s^{3}+(k_s+k_{\upsilon})\omega_{c}s^{2}+\omega_{c}^{2}s+k_{\upsilon}\omega_{c}^{3}}\label{eq_DCSOGI_QSG_def_al}\\
G_{\beta}(s) =\frac{y_{\beta}(s)}{x(s)} =  \frac{k_s\omega_{c}^{2}s}{s^{3}+(k_s+k_{\upsilon})\omega_{c}s^{2}+\omega_{c}^{2}s+k_{\upsilon}\omega_{c}^{3}}\label{eq_DCSOGI_QSG_def_be}\\
G_{\upsilon}(s) =\frac{y_{\upsilon}(s)}{x(s)} =  \frac{k_{\upsilon}\omega_{c}(s^{2}+\omega_{c}^{2})}{s^{3}+(k_s+k_{\upsilon})\omega_{c}s^{2}+\omega_{c}^{2}s+k_{\upsilon}\omega_{c}^{3}}\label{eq_DCSOGI_QSG_def_o}
\end{align}
\end{subequations}
where $k_\upsilon$ refers to the gain of the DC loop\footnote{Note that setting $k_\upsilon = 0$ cancels the DC branch and results in the conventional SOGI-QSG structure and transfer functions.} and $s$ to the complex frequency.
 The evaluation of (\ref{eq_DCSOGI_QSG_def}) at $s = j2\pi f_\gamma$, where $f_\gamma$ represents the frequency of a generic tone '$\gamma$', provides the frequency response of the filter depicted in Fig. \ref{fig_DCSOGI_QSG_Filter_Resp}(a) (Fig. \ref{fig_DCSOGI_QSG_Filter_Resp}(b) corresponds to that of the conventional SOGI i.e. when $k_\upsilon = 0$ is selected) with $\sigma_{\alpha_{\gamma}} = G_{\alpha}(j2\pi f_\gamma),\sigma_{\beta_{\gamma}} = G_{\beta}(j2\pi f_\gamma)$ and $\sigma_{\upsilon_{\gamma}} = G_{\upsilon}(j2\pi f_\gamma)$ representing the magnitude and phase distortion introduced by the filter to each component.
The impacts of the selection of $k_\upsilon$ in the context of a DCSOGI-PLL are discussed in \cite{Karimi-Ghartemani2012_DCSOGI} based on a root locus examination. There, a compromise between large and small values of $k_\upsilon$ is advised with all poles forced to have the same real part presented as an example. Even though the DCSOGI-QSG is not used here as an adaptive filter, the selection of $k_\upsilon$ alters its transfer functions (\ref{eq_DCSOGI_QSG_def}) and, thus, impacts the filter frequency response. A selection of $k_\upsilon$ so that all continuous time (CT) poles share the same real part (based on the values for $k_s$ and $\omega_c$ already selected in \cite{Veloso2024_SOGI_IpDFT_SGSMA,Veloso2023_SOGI_IpDFT_PT}) is here used. This has been found to be an acceptable approach resulting in an adequate SE performance (refer to Section \ref{doc_sec_5} for the performance assessment).
\vspace{-3mm}
\begin{algorithm}
\footnotesize
\caption{DCSOGI-QSG Algorithm}
\label{alg_DCSOGI}
\begin{algorithmic}[1]
\Require $[x(n)]$
\State $\{y_{\alpha}(n)\} = \mathtt{TO\text{-}Int}[(k_s(x(n)-y_{\alpha}(n-1)-y_{\upsilon}(n-1))-y_{\beta}(n-1))\omega_c]$
\State $\{y_{\beta}(n)\} = \mathtt{TO\text{-}Int}[y_{\alpha}(n-1)]\omega_c$
\State $\{y_{\upsilon}(n)\} = \mathtt{TO\text{-}Int}[k_\upsilon \omega_c (x(n)-y_{\alpha}(n-1)-y_{\upsilon}(n-1))]$
\Ensure $\{y_{\alpha}(n),y_{\beta}(n),y_{\upsilon}(n)\}$
\end{algorithmic}
\end{algorithm}

\vspace{-3mm}
As shown by the magnitude response in Fig. \ref{fig_DCSOGI_QSG_Filter_Resp}(a), the DCSOGI rejects the DC offset in both $\alpha$ and $\beta$ components. Moreover, compared to the SOGI (Fig. \ref{fig_DCSOGI_QSG_Filter_Resp}(b)) a very similar response is obtained in terms of $\alpha$ while a higher peak gain is present in the subharmonic region for $\beta$ under the current parametrization. Additionally, while exhibiting similar values, it presents a steeper group delay $(\tau)$ within the range of interest for the fundamental component i.e. $[45-55]$ Hz range $\rightarrow$ $[0.9 - 1.1]$ pu, as shown by the zoomed plots in Fig. \ref{fig_DCSOGI_QSG_Filter_Resp}(a)-(b). The group delay $(\tau)$, also known as envelope delay \cite{Proakis2014_DSPBook}, measures the delay experienced by the envelope of a signal tone '$\gamma$' as it moves through the filer. It is defined as:
\begin{equation}
\label{eq_gd_def}
\tau(\omega) = - \frac{d \Theta (\omega)}{d \omega}
\end{equation} where $\Theta (\omega)$ denotes the phase response of the filter. Given that $\angle\alpha - \angle\beta$ is constant for both SOGI and DCSOGI, both components experience the same group delay $\tau_{\alpha\beta}(\omega)$:
\begin{equation}\label{eq_gd_DCSOGI}
\tau_{\alpha\beta}(\omega) = - \frac{d \angle{\sigma_{\alpha}}(\omega)}{d \omega} = - \frac{d \angle{\sigma_{\beta}}(\omega)}{d \omega}\\ 
\end{equation}
Moreover, both SOGI and DCSOGI have a $\tau_{\alpha\beta}$ which is dependent on the tone's frequency $\omega = 2\pi f$. This means that an adjustment of the analysis window, as detailed in Section \ref{doc_sec_3b}, is necessary for the SE.
As done in \cite{Veloso2023_SOGI_IpDFT_PT,Veloso2024_SOGI_IpDFT_SGSMA} for the SOGI, the discrete time (DT) version of the DCSOGI is obtained using the third-order integrator discretizaion technique, which in \cite{Chibotaru2006_1Ph_SOGI_PLL} was shown for the former to offer the best results among several discretization techniques. It relates $s$ and $z$ domains according to:
\begin{equation}\label{eq_TO_Disct_def}
\frac{1}{s} = \frac{T_s}{12} \frac{23 z^{-1} -16z^{-2} +5z^{-3}}{1-z^{-1}}
\end{equation}
where $T_s$ denotes the sampling time. A straight implementation of the filter is done through Algorithm \ref{alg_DCSOGI} which is the result of applying (\ref{eq_TO_Disct_def}) to Fig. \ref{fig_DCSOGI_QSG_FlowChart} and where $\mathtt{TO\text{-}Int}$ refers to \cite[Algorithm 2]{Veloso2023_SOGI_IpDFT_PT}. \vspace{-3mm}
\section{The DCSOGI-IpDFT Algorithm}\label{doc_sec_3}
\subsection{Auxiliary DC Blocker Filter}\label{doc_sec_3a}
The SOGI-IpDFT \cite{Veloso2024_SOGI_IpDFT_SGSMA,Veloso2023_SOGI_IpDFT_PT} relied on the use of the original unfiltered signal $x(n)$ spectrum for OOBI detection. This is because, due to the employed detection mechanism, the asymmetric magnitude gains around $f_n$ of the SOGI filter difficult the unequivocal identification of an OOBI. While the DCSOGI allows for the cancellation of the DC offsets present in $\alpha\beta$, the filter also exhibits an asymmetric magnitude response around $f_n$. Thus, in order for the derived SE technique to be fully resilient against DC offsets, a DC free signal must be used for the OOBI detection. 

\begin{figure}[!t]
\centering
\includegraphics[width=\linewidth]{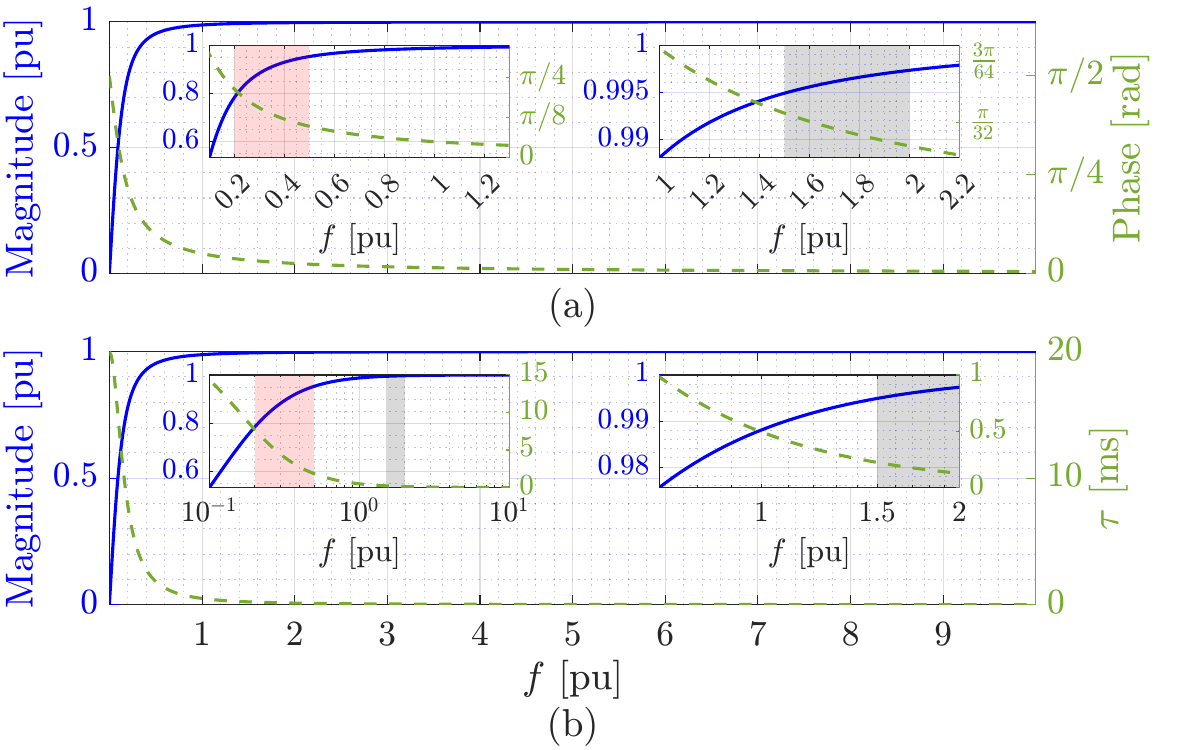}
\vspace{-5mm}
\caption{Frequency Response of the DC Blocker filter for a normalized frequency $f\text{[pu]} = f/f_n$: (a) Magnitude and Phase response; (b) Magnitude Response and Group Delay. Zoomed plots are provided for further details around the OOBI frequencies with shaded areas corresponding to the subharmonic (red) and interharmonic (gray) ranges.}
\label{fig_DC_Block_FR}
\vspace{-5mm}
\end{figure}

To attain this, a DC Blocker filter is employed to eliminate the DC content from $x(n)$ without compromising the spectral energy dispersion around $f_n$. The non-linear recursive IIR DC Blocker implementation presented in \cite{Yates2008_DC_Blocker} is here adopted due to its simplicity. It is defined by the following difference equation and transfer function:
\begin{subequations}\label{eq_IIR_DC_Block_def}
\begin{align}
x_{\varnothing}(n) &= x(n) - x(n-1) + p \ x_{\varnothing}(n-1) \label{eq_IIR_DC_Block_diff_eq} \\ 
H_{\varnothing}(z) &= \frac{1-z^{-1}}{1-pz^{-1}} \label{eq_IIR_DC_Block_TF}
\end{align}
\end{subequations}
where $x_\varnothing(n)$ denotes the filtered output, $x(n)$ the input signal and $p$ the filter's pole. A value of $p=0.999<1$ is chosen as a compromise between the steepness of the transition band\footnote{Note that a steeper transition band translates into: (i) a flatter and more symmetrical passband around $f_n$ and (ii) larger group delay values for those frequencies closest to DC.} and the filter's responsiveness at low frequencies. The filter's frequency response ($\sigma_{\varnothing}(\omega) = H_{\varnothing}(z=e^{j\omega T_s})$) is shown in Fig. \ref{fig_DC_Block_FR} for a normalized frequency. Zoomed plots are provided for further detail within the OOBI range of interest showing how the selected $p$ does not significantly attenuate the lowest OOBI frequency allowing its proper identification (see Section \ref{doc_sec_4}). As shown, the filter presents a variable $\tau_{\varnothing}(\omega)$ with rapidly decaying values as the frequency of the tone increases\footnote{Please note that, while $\tau(\omega)$ has units of time, for the remainder of this work we will refer to all $\tau(\omega)$ as their rounded to the nearest integer sample. Thus, hereafter any $\tau$ is defined as $\tau(\omega) := \left\lfloor- \frac{d \Theta (\omega)}{d \omega} \frac{1}{T_s} \right\rceil$.}. \vspace{-3mm}
\subsection{Two and Three Nominal Cycle Variants}\label{doc_sec_3b}
Both the two and three-cycle variants of the DCSOGI-IpDFT (henceforth referred to as '2c' and '3c') consider the same static signal model presented in \cite{Veloso2023_SOGI_IpDFT_PT,Veloso2024_SOGI_IpDFT_SGSMA}, i.e., the sequence of $N$ samples within the analysis window of length $T = NT_s$ is the superposition of a fundamental and a potential interference tone, each defined by $\{A_0,f_0,\varphi_0\}$ and $\{A_i,f_i,\varphi_i\}$, which refer to their amplitude, frequency and initial phase. The pseudocode in Algorithm \ref{alg_2c3c_DCSOGI_IpDFT} summarizes the steps required by both the '3c' and '2c' variants to provide an estimate of the signal parameters at the reporting instant $n_r = \kappa\frac{F_s}{F_r} + N-1 -\tau_w; \; \forall \kappa \in \mathbb{Z}$, where $F_s$ and $F_r$ correspond respectively to the sampling and reporting rates and $\tau_w$ to the delay introduced by the observation window. The exclusive steps for each variant are highlighted in different colors (\textcolor{ForestGreen}{'3c'},\textcolor{orange}{'2c'}).

\begin{figure}[!t]
\centering
\includegraphics[width=\linewidth]{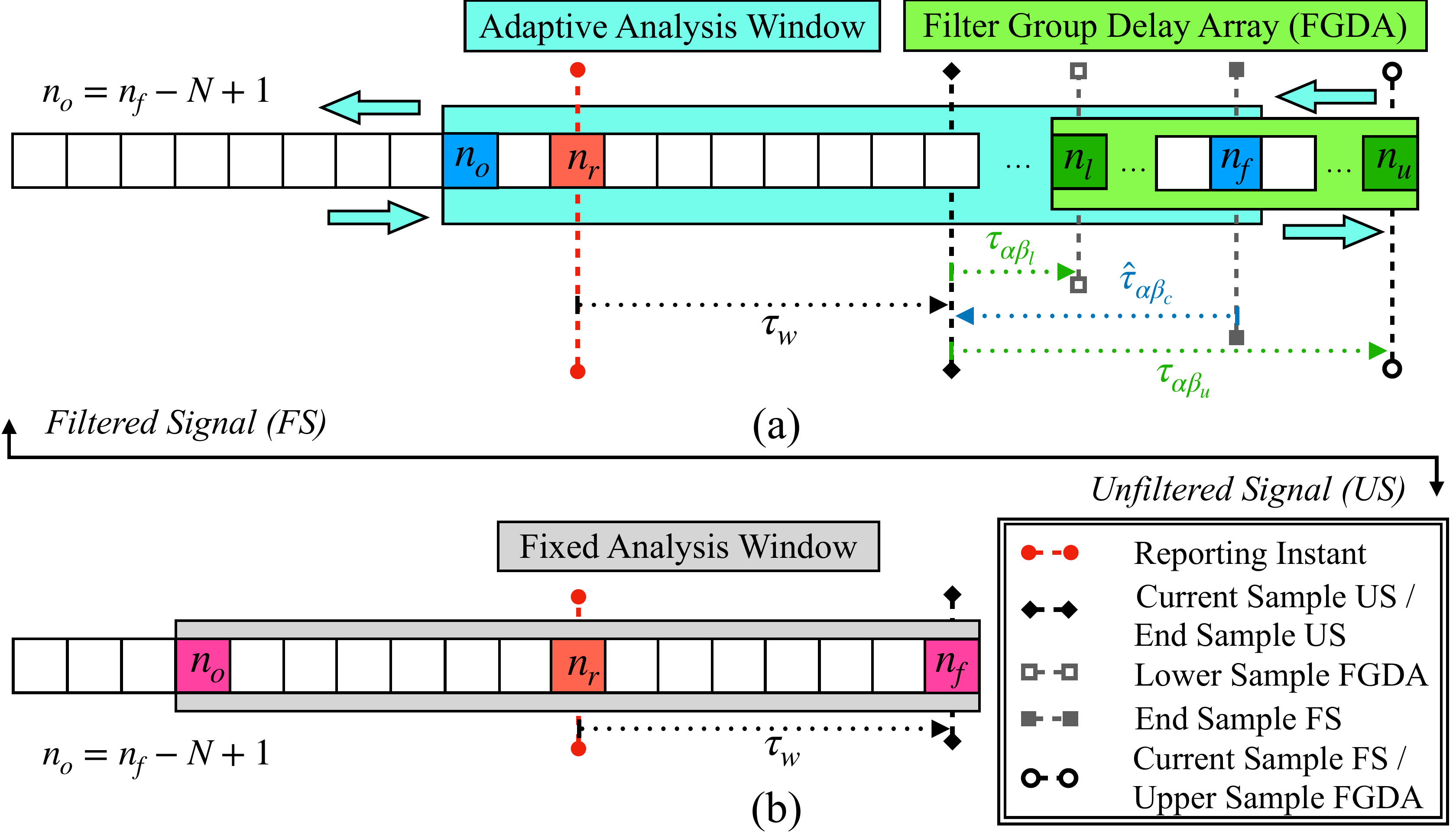}
\vspace{-5mm}
\caption{Window selection procedure to ensure a constant reporting rate $F_r$ is maintained in the cases of: (a) a signal filtered by a variable group delay filter and (b) an unfiltered signal.}
\label{fig_GD_Adaptive_window}
\vspace{-5mm}
\end{figure}

First, $y_{\alpha}(n)$ and $y_{\beta}(n)$, resulting from the filtering of $x(n)$ through the DCSOGI-QSG (line 1), are obtained. For techniques that operate directly on the unfiltered signal $x(n)$, such as the e-IpDFT\cite{Romano2014EnhancedInterpolatedDFT}, i-IpDFT\cite{Derviskadic2018i-IpDFTforSynchrophasorEstimation,Derviskadic2020_iIpDFT_FPGA} or FiIpDFT\cite{Song2022_FiIpDFT,Song2022_FiIpDFT_Addendum}, or for those where only filters with a generalized linear phase \cite{OppenheimAlan2013DSPP} response are employed, the observation window can be accounted for (and compensated) in a straightforward manner. Indeed, all delays present (i.e. those introduced by the analysis window and the filters used) are constant and known a priori. However, this is not the case for both the DCSOGI-QSG and the DC-Blocker filters, as their group delays $\tau(\omega)$ are a function of the tone's frequency. In turn, this means that a repositioning of the analysis window based on the signal frequency might be necessary to ensure that the calculated estimates $\hat{\mathcal{S}}_{0}(n_r)=\{\hat{f}_{0}(n_r),\hat{A}_{0}(n_r),\hat{\varphi}_{0}(n_r)\}$ correspond to the required reporting time $n_r$. The procedure is illustrated in Fig. \ref{fig_GD_Adaptive_window}(a) for the case of the DSOCGI-QSG, (the same technique is adopted for the DC-Blocker) and compared with that used in methods that operate directly on $x(n)$ (Fig. \ref{fig_GD_Adaptive_window}(b)). The end sample of the analysis window $n_f$ is selected within the array\footnote{ $\tau_{\alpha\beta_l}$ and $\tau_{\alpha\beta_u}$ denote the DCSOGI-QSG group delays, respectively, at 55 and 45 Hz. Notice that, as shown in Fig. \ref{fig_DCSOGI_QSG_Filter_Resp}(a), the group delay is monotonically decreasing within such range, i.e. the largest group delay corresponds to the smallest frequency and vice versa.} $[n_l ,n_u ]$ ($n_l = \kappa\frac{F_s}{F_r} +N-1 + \tau_{\alpha\beta_l}; \;n_u = \kappa\frac{F_s}{F_r} +N-1 + \tau_{\alpha\beta_u}$) so that all potential group delays introduced by the filter in the [45-55] Hz range can be accounted for\footnote{The $f_n \pm 5$ Hz range is considered as per M-class requirements in \cite{PMU_Measurement_60255-118-1-2018} while for the DC-Blocker, due to its smaller delay values around the fundamental frequency, a larger $f_n \pm 10$ Hz range has been considered.}. $[n_l ,n_u ]$ is referred to as 'Filter Group Delay Array' within Fig. \ref{fig_GD_Adaptive_window}(a). 

An initial window, considering nominal system frequency $f_n$, is selected (lines 2-3), where $n_o$ represents the initial sample within the analysis window, $\tau_{\alpha\beta_{n}}$ the DCSOGI-QSG group delay at $f_n$ and $\tau_{y_{\alpha\beta_o}}$ the distance between $n_u$ and $n_f$ for this initial window. The DFT spectrums of $y_{\alpha}(n)$ and $y_{\beta}(n)$ are then obtained and windowed in the frequency domain $(Y_{\alpha_{H}}(k),Y_{\beta_{H}}(k))$ (lines 4-5). A first frequency estimate $(\hat{f}_0)$ is then obtained by applying an IpDFT to the $Y_{\alpha_{H}}(k)+jY_{\beta_{H}}(k)$ spectrum (line 6) and used to characterize: (i) both the magnitude and phase distortion introduced by the filter $(\sigma_{\alpha_{0}},\sigma_{\beta_{0}})$, depicted in Fig. \ref{fig_DCSOGI_QSG_Filter_Resp}(a), on the fundamental component by means of (\ref{eq_DCSOGI_QSG_def_al})-(\ref{eq_DCSOGI_QSG_def_be}) (line 6) as well as (ii) the group delays introduced by both filters ($\hat{\tau}_{\alpha\beta_{c}}$ and $\hat{\tau}_{\varnothing_{c}}$) (line 7). The magnitude distortion will later be removed from the $\alpha$ and $\beta$ spectrums ($(Y_{\alpha_{H}}(k)/|\sigma_{\alpha_{0}}^{q-1}|, Y_{\beta_{H}}(k)/|\sigma_{\beta_{0}}^{q-1}|)$) so that $\xi \simeq 1$ (line 16). With $\hat{\tau}_{\alpha\beta_{c}}$ a refinement of the analysis window is done and $Y_{\alpha_{H}}(k)$ and $Y_{\beta_{H}}(k)$ recalculated (lines 8-9). Likewise, the DC-Blocker filtered signal $x_{\varnothing}(n)$ is obtained and its windowed DFT spectrum ($X_{\varnothing_{H}}(k)$) calculated using $\hat{\tau}_{\varnothing_{c}}$ to correctly place its analysis window (lines 10-13). Given the difference between the group delays of both filters (DCSOGI-QSG and DC-Blocker), a correction of the offset $d_\varnothing$ between their respective maximum delays ($\tau_{\alpha\beta_u}$ and $\tau_{\varnothing_u}$) is required to align $x_{\varnothing}(n)$ with $y_{\alpha}(n)$ and $y_{\beta}(n)$. Lastly, an iterative compensation loop (lines 15-43) is executed to detect and, if so, remove the effects of a potential OOBI or 2\textsuperscript{nd} harmonic interference. Previously, the initial values of the DCSOGI-QSG filtered spectrum of this interference $\hat{Y}_{i}^{0}(k)$ and its trigger detection flag $\eta_{i}$ are set to 0 (line 14).

Inside, the estimates of the fundamental tone $(\hat{f}_{0}^{q},\hat{A}_{0}^{q},\hat{\varphi}_{0}^{q})$ are given by a second IpDFT applied after removing the estimated contribution of the DCSOGI-QSG filtered interference tone $(\hat{Y}_{i}^{q-1}(k))$ from the magnitude corrected spectrum $Y_{\alpha_{H}}(k)/|\sigma_{\alpha_{0}}^{q-1}| + jY_{\beta_{H}}(k)/|\sigma_{\beta_{0}}^{q-1}|$ (line 16). These are used in the first iteration ($q = 1$) to assess the existence of an interference (lines 18-26). In contrast to \cite{Veloso2024_SOGI_IpDFT_SGSMA,Veloso2023_SOGI_IpDFT_PT}, and to avoid potential false triggers caused by the presence of DC offsets, this detection is now done based on the calculation of the residual spectrum of the DC-Blocker-filtered signal $x_\varnothing(n)$ instead of that of the original input $x(n)$. This residual spectrum $\hat{X}_{\varnothing_{i}}(k)$ is obtained by removing that of the estimated fundamental $\hat{X}_{\varnothing_{0}}(k)$ from the original $X_{\varnothing_{H}}(k)$ (line 22). $\hat{X}_{\varnothing_{0}}(k)$ is calculated through $\mathtt{wf}$ using the $\varnothing$ estimated parameters of the fundamental ($|\sigma_\varnothing|\hat{A}_{0}$ and $\hat{\varphi}_{0_{\varnothing}}$)\footnote{The $\varnothing$ magnitude $|\sigma_\varnothing|\hat{A}_0$ is obtained by considering the gain introduced by the DC-Blocker $|\sigma_\varnothing|$. The $\varnothing$ phase $\hat{\varphi}_{0_{\varnothing}}$ results from the removal of the phase shift introduced by the DCSOGI-QSG filter $\angle{\sigma_{\alpha_{0}}}$, the addition of the corresponding phase distortion caused by the DC-Blocker $\angle\sigma_\varnothing$ and the compensation of the phase shift ($2\pi\hat{f}_0(\hat{\tau}_{\alpha\beta_c}-\hat{\tau}_{\varnothing_c})$) caused by the time offset between the analysis windows of both filtered signals.} (line 21), both determined taking into account the DCSOGI-QSG and DC-Blocker filters response (lines 19-20).

\centerline{\begin{minipage}[!t]{\linewidth}
\begin{algorithm}[H]
\footnotesize
\caption{DCSOGI-IpDFT: \textcolor{ForestGreen}{3 cycle} and \textcolor{orange}{2 cycle} variants}
\label{alg_2c3c_DCSOGI_IpDFT}
\begin{algorithmic}[1] \setstretch{1.35}
\Require $[x(n)]; \; n \in [\kappa\frac{F_s}{F_r}, \kappa\frac{F_s}{F_r} +N-1 + \tau_{\alpha\beta_u}];\; \forall \kappa \in \mathbb{Z}$
\State $\{y_{\alpha}(n),y_{\beta}(n)\} = \mathtt{DCSOGI\text{-}QSG}[x(n)]$
\State $n_f = \kappa\frac{F_s}{F_r} +N-1 + \tau_{\alpha\beta_u}-\tau_{y_{\alpha\beta_o}};$  where $\tau_{y_{\alpha\beta_o}} = \tau_{\alpha\beta_u} - \tau_{\alpha\beta_{n}}$
\State $n_o = n_f - N +1$
\State $Y_{\{\alpha,\beta\}}(k) = \mathtt{DFT}[{y}_{\{\alpha,\beta\}}(n)]; \; n \in [n_o, n_f]$ 
\State $ Y_{\{\alpha,\beta\}_{H}}(k)  = \mathtt{Hann}[Y_{\{\alpha,\beta\}}(k)]$
\State $\{\hat{f}_{0}\} = \mathtt{IpDFT}[Y_{\alpha_{H}}(k) + j Y_{\beta_{H}}(k)]$; $\{\sigma_{\alpha_{0}}^{0},\sigma_{\beta_{0}}^{0}\}=\mathtt{DCSOGI\text{-}CG_{\alpha\beta}}[\hat{f}_{0}]$
\State $\{\hat{\tau}_{\alpha\beta_{c}}\} = \mathtt{DCSOGI\text{-}\tau}[2\pi\hat{f}_{0}]$;  $\{\hat{\tau}_{\varnothing_{c}}\} = \mathtt{DCBlocker\text{-}\tau}[2\pi\hat{f}_{0}]$ 
\State $n_f = \kappa\frac{F_s}{F_r} +N-1 + \tau_{\alpha\beta_u}-\tau_{y_{\alpha\beta}}$; where $\tau_{y_{\alpha\beta}} = \tau_{\alpha\beta_u} - \hat{\tau}_{\alpha\beta_{c}}$
\State Apply lines 3-5
\State $x_{\varnothing}(n) = \mathtt{DC\text{-}Block}[x(n))]$;
\State $n_f = \kappa\frac{F_s}{F_r} +N-1 + \tau_{\alpha\beta_u}-(d_{\varnothing}+\tau_{x_{\varnothing}});$  where $\tau_{x_{\varnothing}} = \tau_{\varnothing_u} - \hat{\tau}_{\varnothing_{c}}$
\State $n_o = n_f - N +1$
\State $X_{\varnothing}(k) = \mathtt{DFT}[x_{\varnothing}(n)]$; $\; n \in [n_o, n_f]$; $ X_{\varnothing_{H}}(k)  =  \mathtt{Hann}[X_{\varnothing}(k)]$
\State $\textbf{Initialization: }{\hat{Y}_{i}^{0}(k)} = 0; \eta_{i} = 0$
\For{$q=1$ \textbf{to} $Q$}
    \State $\{\hat{f}_{0}^{q},\hat{A}_{0}^{q},\hat{\varphi}_{0}^{q}\} =\mathtt{IpDFT}\left[\frac{Y_{\alpha_{H}}(k)}{|\sigma_{\alpha_{0}}^{q-1}|}+j\frac{Y_{\beta_{H}}(k)}{|\sigma_{\beta_{0}}^{q-1}|} - \hat{Y}_{i}^{q-1}(k) \right]$
    \State $\hat{A}_{0}^{q} = \hat{A}_{0}^{q}/2$; $\{\sigma_{\alpha_{0}}^{q},\sigma_{\beta_{0}}^{q}\}=\mathtt{DCSOGI\text{-}CG_{\alpha\beta}}[\hat{f}_{0}^{q}]$
    \If{$q = 1$}
    \State $\{\sigma_{\varnothing}\} = \mathtt{DCBlocker\text{-}CG}[\hat{f}_{0}]$
    \State $\hat{\varphi}_{0_{\varnothing}}=\hat{\varphi}_{0}-\angle{\sigma_{\alpha_{0}}}+\angle{\sigma_{\varnothing}}-2\pi\hat{f}_{0}(\hat{\tau}_{\alpha\beta_{c}}-\hat{\tau}_{\varnothing_{c}})$
        \State $\hat{X}_{\varnothing_{0}}(k) = \mathtt{wf}[\hat{f}_{0},\hat{A}_{0}|\sigma_{\varnothing}|,\hat{\varphi}_{0_{\varnothing}}]+\mathtt{wf}[\text{--}\hat{f}_{0},\hat{A}_{0}|\sigma_{\varnothing}|,\text{--}\hat{\varphi}_{0_{\varnothing}}]$        
        \State $\hat{X}_{\varnothing_{i}}(k) = X_{\varnothing_{H}}(k) - \hat{X}_{\varnothing_{0}}(k)$
        \State Apply (\ref{eq_Eo_def})\textcolor{ForestGreen}{-(\ref{eq_kc_range_def})-(\ref{eq_Ec_def})}\textcolor{orange}{-(\ref{eq_Eint_def})-(\ref{eq_yc_def})-(\ref{eq_env_def})-(\ref{eq_ang_def})}
        \If{$\frac{\{\textcolor{ForestGreen}{E_{c}},\textcolor{orange}{E_{i}}\}}{E_{o}} \text{$>$}\text{$\lambda$} $ \textcolor{orange}{\textbf{\&} Var$(y_{\phi}(n)) \text{$>$}\text{$\lambda_{\phi}$}$ \textbf{\&} Var$({y_{A}(n)})/\hat{A}_{0}^2 \text{$>$}\text{$\lambda_{A}$}$}} 
            \State $\eta_{i} = 1$; \textbf{else }\text{Apply (\ref{eq_E2_def})} 
            \textcolor{orange}{\IIf{$\frac{E_{int}}{E_o}\text{$>$}\text{$\lambda_{int}$}$} $k_i = k_{int}$ \ElseIIf  $k_i = k_{sub}$}
        \EndIf
    \EndIf
    \If{$\eta_{i} = 1$}
    \State $\hat{A}_{0_{\beta}}^{q}=\hat{A}_{0}^{q}|\sigma_{\beta_{0}}^{q}|;\hat{\varphi}_{0_{\beta}}^{q}=\hat{\varphi}_{0}^{q}-\pi/2$
    \State ${\hat{Y}_{0_{\beta}}^{q}(k)} = \mathtt{wf}[\hat{f}_{0}^{q},\hat{A}_{0_{\beta}}^{q},\hat{\varphi}_{0_{\beta}}^{q}] + \mathtt{wf}[\text{--}\hat{f}_{0}^{q},\hat{A}_{0_{\beta}}^{q},\text{--}\hat{\varphi}_{0_{\beta}}^{q}]$
    \State $\{\hat{f}_{i}^{q},\hat{A}_{i_{\beta}}^{q},\hat{\varphi}_{i_{\beta}}^{q}\} = \mathtt{e\text{-}IpDFT}[Y_{\beta_{H}}(k) -{\hat{Y}_{0_{\beta}}^{q}(k)}]\textcolor{orange}{|_{k_i}}$
    \State $\{\sigma_{\alpha_{i}}^{q},\sigma_{\beta_{i}}^{q}\}=\mathtt{DCSOGI\text{-}CG_{\alpha\beta}}[\hat{f}_{i}^{q}]$
    \State $\hat{A}_{i}^{q}= \hat{A}_{i_{\beta}}^{q}/|\sigma_{\beta_{i}}^{q}|;\hat{\varphi}_{i}^{q} = \hat{\varphi}_{i_{\beta}}^{q} - \angle{\sigma_{\beta_{i}}^{q}}$
    \State $\{\sigma_{\text{+}}^{q},\sigma_{\text{--}}^{q}\} = \mathtt{DCSOGI\text{-}CG_{+-}}[\sigma_{\alpha_{i}}^{q},\sigma_{\beta_{i}}^{q},\sigma_{\alpha_{0}}^{q},\sigma_{\beta_{0}}^{q}]$
    \State $\hat{A}_{i\text{+}}^{q}=\hat{A}_{i}^{q}|\sigma_{\text{+}}^{q}|;\hat{A}_{i\text{--}}^{q}=\hat{A}_{i}^{q}|\sigma_{\text{--}}^{q}|$
    \State $\hat{\varphi}_{i\text{+}}^{q}=\hat{\varphi}_{i}^{q}+\angle{\sigma_{\text{+}}^{q}};\hat{\varphi}_{i\text{--}}^{q}=-\hat{\varphi}_{i}^{q}+\angle{\sigma_{\text{--}}^{q}}$
    \State ${\hat{Y}_{i}^{q}(k)} = \mathtt{wf}[\hat{f}_{i}^{q},\hat{A}_{i\text{+}}^{q},\hat{\varphi}_{i\text{+}}^{q}] + \mathtt{wf}[\text{--}\hat{f}_{i}^{q},\hat{A}_{i\text{--}}^{q},\hat{\varphi}_{i\text{--}}^{q}]$
    \ElsIf{$E_{2}/E_{i}$ \text{$>$}\text{$\lambda_2$} }
        \State Lines 28 - 29 \textcolor{orange}{(Applied to the $\alpha$ spectrum)}
        \textcolor{orange}{\State{$\{\hat{f}_{i}^{q},\hat{A}_{i_{\alpha}}^{q},\hat{\varphi}_{i_{\alpha}}^{q}\} = \mathtt{e\text{-}IpDFT}[Y_{\alpha_{H}}(k) -\hat{Y}_{0_{\alpha}}^{q}(k)]|_{k_i = 4}$}}
        \textcolor{ForestGreen}{\State $\{\hat{f}_{i}^{q},\hat{A}_{i_{\beta}}^{q},\hat{\varphi}_{i_{\beta}}^{q}\} = \mathtt{e\text{-}IpDFT}[Y_{\beta_{H}}(k) -{\hat{Y}_{0_{\beta}}^{q}(k)}]|_{k_i = 6}$}
        \State Lines 31 - 36 \textcolor{orange}{(Applied to the $\alpha$ spectrum)}
    \Else 
        \State \textbf{break}
    \EndIf
\EndFor
\State $\hat{\varphi}_{0}(n_r) = \hat{\varphi}_{0}^{q} - \angle{\sigma_{\alpha_{0}}^{q}} + \frac{2\pi\hat{f}_0}{F_s}(\frac{N}{2}-\hat{\tau}_{\alpha\beta_c});\,\hat{f}_{0}(n_r)\text{=} \hat{f}_{0}^{q};\,\hat{A}_{0}(n_r)\text{=}\hat{A}_{0}^{q}$
\Ensure $\{\hat{f}_{0}(n_r),\hat{A}_{0}(n_r),\hat{\varphi}_{0}(n_r)\};$ $n_r = \kappa\frac{F_s}{F_r} + N-1 -\tau_w; \; \forall \kappa \in \mathbb{Z}$
\end{algorithmic}
\end{algorithm}\vspace{-5.5mm}\footnotetext{$\mathtt{DCSOGI\text{-}QSG}$ corresponds to Algorithm \ref{alg_DCSOGI}, $\mathtt{DCSOGI\text{-}CG_{\alpha\beta}}$ to (\ref{eq_DCSOGI_QSG_def_al})-(\ref{eq_DCSOGI_QSG_def_be}) evaluated at $s = j2\pi f$, $\mathtt{DCSOGI\text{-}\tau}$ and  $\mathtt{DCBlocker\text{-}\tau}$ to the round to the nearest integer discrete approximations of (\ref{eq_gd_def}) evaluated for each respective filter, $\mathtt{DC\text{-}Block}$ to (\ref{eq_IIR_DC_Block_diff_eq}), $\mathtt{DCBlocker\text{-}CG}$ to (\ref{eq_IIR_DC_Block_TF}) evaluated at $z = e^{j2\pi f/F_s}$, $\mathtt{DCSOGI\text{-}CG_{+-}}$ to \cite[Algorithm 3]{Veloso2023_SOGI_IpDFT_PT} and \cite[Algorithm 4]{Veloso2024_SOGI_IpDFT_SGSMA}, $\mathtt{e\text{-}IpDFT}$ to \cite[Algorithm 1]{Derviskadic2018i-IpDFTforSynchrophasorEstimation}, $\mathtt{IpDFT}$ to \cite[eq.(2)-(3)]{Veloso2024_SOGI_IpDFT_SGSMA}, $\mathtt{wf}$ to $\hat{X}_{H\pm}(k) =  \hat{A} e^{\pm j\hat{\varphi}} W_{H}(k\mp\hat{f}T)$, where $W_{H}(k)$ is the DFT of the Hanning window, and $\mathtt{Hann}$ to $X_{H}(k) =  0.5X(k) -0.25(X(k-1) + X(k+1))$ i.e. its application in the frequency domain.}
\end{minipage}}


Unlike \cite{Veloso2024_SOGI_IpDFT_SGSMA,Veloso2023_SOGI_IpDFT_PT}, where the '3c' variant relied on the technique proposed in \cite{Derviskadic2018i-IpDFTforSynchrophasorEstimation} to verify the existence of an OOBI tone, the same refined bin selection criterion proposed in \cite{Veloso2023_TIM} is now adopted. In \cite{Derviskadic2018i-IpDFTforSynchrophasorEstimation} the ratio between the energy contained in the complete residual $(E_i)$ and original $(E_o)$ (\ref{eq_Eo_def}) spectra was used and compared to a threshold level tuned heuristically $(\lambda)$ to determine the presence of an OOBI\footnote{Please note that those expressions equivalent to (\ref{eq_Eo_def}) in \cite{Derviskadic2018i-IpDFTforSynchrophasorEstimation,Veloso2023_SOGI_IpDFT_PT} and (\ref{eq_Eo_def})-(\ref{eq_Eint_def}) in \cite{,Veloso2024_SOGI_IpDFT_SGSMA} refer to the residual and original spectra of the unfiltered input $x(n)$. Here (\ref{eq_Eo_def})-(\ref{eq_Eint_def}) refer, instead, to those of the filtered signal $x_\varnothing(n)$.}. However, now only the energy concentrated in a series of bins of the residual spectrum $(E_c)$ (\ref{eq_Ec_def}) is weighted against $E_o$ (line 24). These bins correspond to the highest magnitude bin $k_c$ among those where a potential OOBI is expected (\ref{eq_kc_range_def}) and its two closest neighbors. 
\begin{subequations}\label{eq_Es_def}
\begin{align}
E_{\{o,i\}} &=  \sum_{k=0}^{K-1}|\{X_{\varnothing_{H}},\hat{X}_{\varnothing_{i}}\}(k)|^{2}\label{eq_Eo_def}\\
E_{\{int,sub\}} &=  |\hat{X}_{\varnothing_{i}}(\{4,0\})|^{2} \label{eq_Eint_def} \\
E_{2} &=  \sum_{k}|\hat{X}_{\varnothing_{i}}|^{2}\;
\begin{cases}
k \in [6, 7], & \text{if} \;\; T = 3/f_n \\
k = 4, & \text{if} \;\; T = 2/f_n \end{cases} \label{eq_E2_def} \\
k_{c} &=  \arg \max_{k}{|\hat{X}_{\varnothing_{i}}|}; \quad k \in [0, 2] \cup [4, 7] \label{eq_kc_range_def} \\
E_{c} &=  \sum_{k}|\hat{X}_{\varnothing_{i}}|^{2}; \;
\begin{cases}
k \in [0, 2], & \text{if } k_{c} = 0 \\
k \in [5, 7], & \text{if } k_{c} = 7 \\
k \in [k_{c}\pm 1], & \text{otherwise} \end{cases} \label{eq_Ec_def}
\end{align}
\end{subequations}
The '2c' preserves the same detection mechanism presented in \cite{Veloso2024_SOGI_IpDFT_SGSMA} i.e. it considers the ratio between $E_i$ and $E_o$ (\ref{eq_Eo_def}) but on a reduced number of DFT bins $K$ (see Section \ref{doc_sec_4}) as well as two additional metrics based on the magnitude-corrected signal $\bar{y}_{c}(n)$ (\ref{eq_yc_def}). These are the variances of: (i) its envelope normalized $\text{Var}(y_{A}(n))/\hat{A}_0^2$ (\ref{eq_env_def}) and (ii) the finite differences of its unwrapped phase $\text{Var}(y_{\phi}(n))$ (\ref{eq_ang_def}).
\begin{subequations}\label{eq_env_ang_def}
\begin{align}
\bar{y}_{c}(n) &=  y_\alpha (n)/|\sigma_{\alpha_{0}}|+ j y_\beta (n)/|\sigma_{\beta_{0}}|; \, n \in [0, N-1] \label{eq_yc_def}\\
y_{A}(n) &=  |\bar{y}_{c}(n)|; \qquad\qquad\qquad\qquad\;\;\; n \in [0, N-1] \label{eq_env_def}\\
y_{\phi}(n) &=  \angle \bar{y}_{c}(n)_{|[\rightleftharpoons]} - \angle \bar{y}_{c}(n\text{ --}1)_{|[\rightleftharpoons]}; \, n \in [1, N-1]\label{eq_ang_def}
\end{align}
\end{subequations}where $\angle \bar{y}_{c}(n)_{|[\rightleftharpoons]}$ represents the unwrapped phase of the complex signal $\bar{y}_{c}(n)$ and ${\mathcal{U}_\mathcal{N}}[.]$ denotes the angle unwrap function so that $\angle \bar{y}_{c}(n)_{|[\rightleftharpoons]} = {\mathcal{U}_\mathcal{N}}[\angle \bar{y}_{c}(n)_{|[-\pi,\pi]}]$, with $\angle \bar{y_c}(n)_{|[-\pi,\pi]}$ representing the wrapped phase of $\bar{y}_{c}(n)$ within $[-\pi,\pi]$.
For the '2c' these additional metrics must be compared with their respective heuristically tuned thresholds $(\lambda_A$ and $\lambda_\phi)$ (Section \ref{doc_sec_4}) to accurately detect the presence of an OOBI (line 24).

\begin{figure}[t]
\centering
\includegraphics[width=\linewidth]{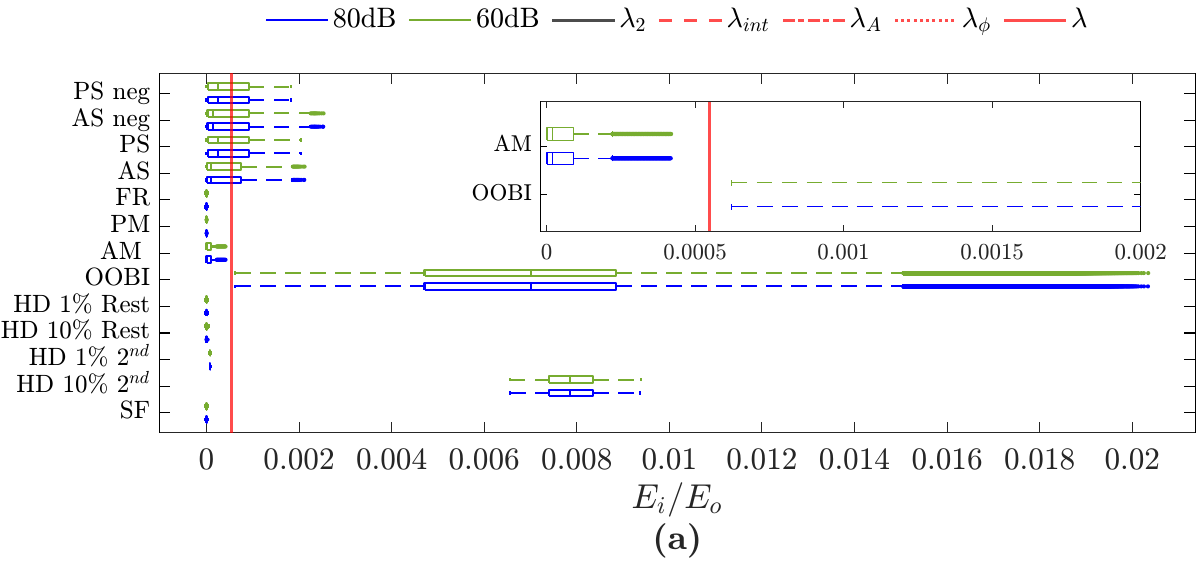}
\includegraphics[width=\linewidth]{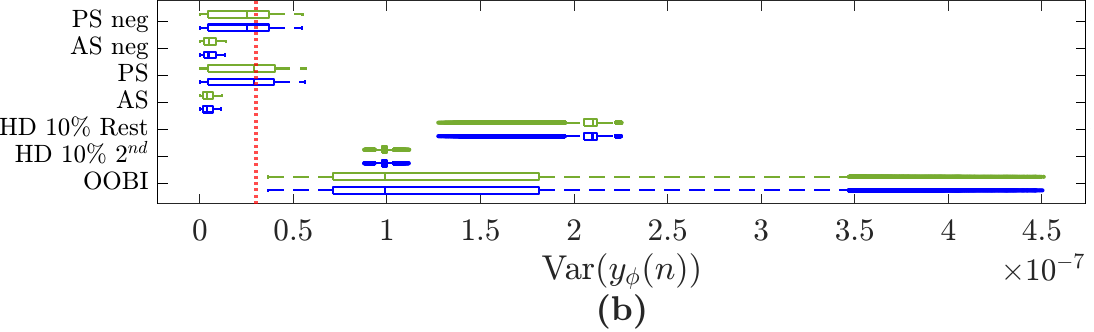}
\includegraphics[width=1.1\linewidth]{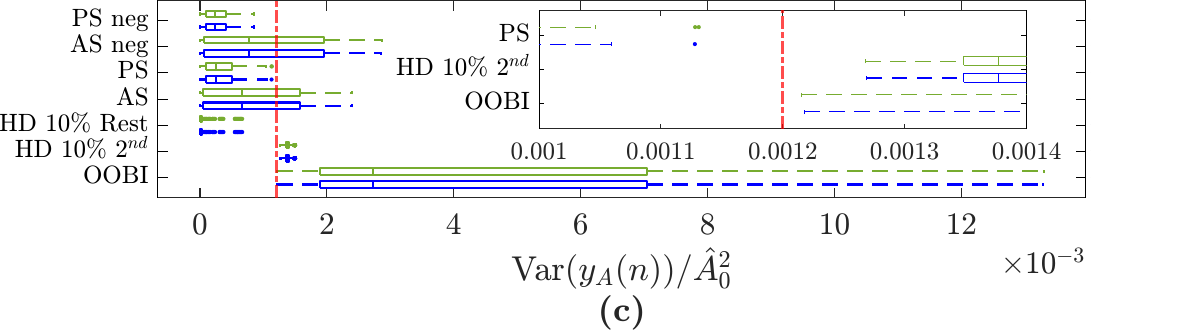}
\includegraphics[width=\linewidth]{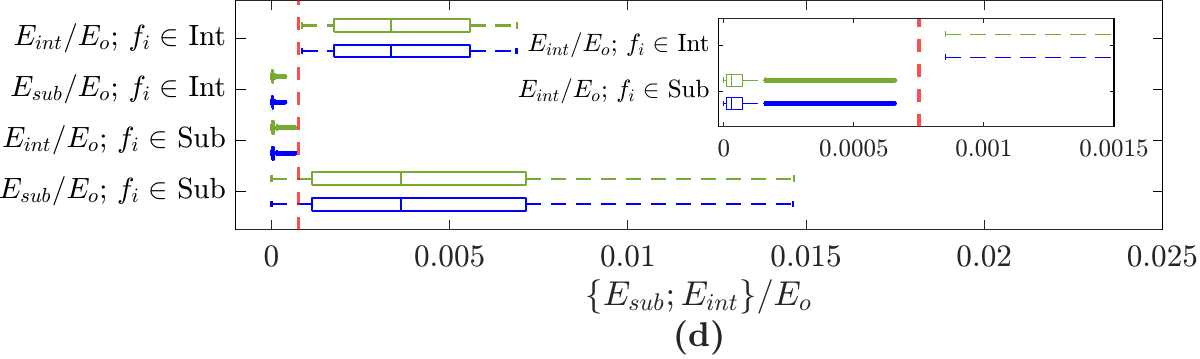}
\includegraphics[width=\linewidth]{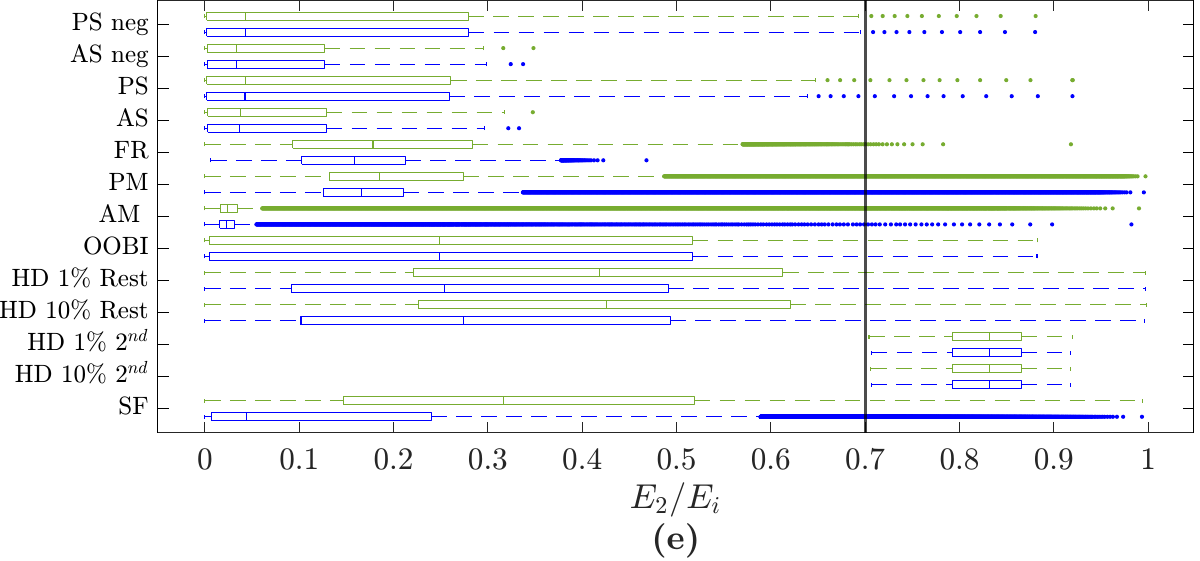}
\vspace{-5mm}
\caption{Boxplot representation for the '2c' algorithm under 80 and 60 dB AWGN  of: (a) $E_i / E_o$ in all testing conditions in \cite{PMU_Measurement_60255-118-1-2018}, (b) Var$(y_{\phi}(n))$ and (c)Var$(y_{A}(n))/\hat{A}_0^2$ during the step, OOBI and HD 10\% tests, (d) $E_{int}/E_o$ and $E_{sub}/E_o$ respectively in the presence of $10 \%$ interharmonic and subharmonic tones and (e) $E_2/E_i$ again in all testing conditions in \cite{PMU_Measurement_60255-118-1-2018}.}
\label{fig_OOBI_threshold_2c}
\vspace{-5mm}
\end{figure}

If an OOBI is detected $(\eta_i=1)$, the '2c' requires an additional check to determine if it falls within the sub- or interharmonic range (line 26). This is needed because the highest bin within the entire OOBI range ceases to be a reliable method to precisely locate the interference when the smaller observation window is adopted. Instead, the energy ratio of the last bin within the interharmonic range $k = 4$ $(E_{int})$ (\ref{eq_Eint_def}) and $E_o$ is compared to the threshold $\lambda_{int}$, so the that appropriate bin can be located according to:
\begin{equation}\label{eq_k_intsub_def}
k_{\{sub,int\}} =  \arg \max_{k}{|\hat{X}_{i}(k)|}; \underbracket[.5pt]{k \in [0, 1]}_{sub}; \underbracket[.5pt]{k \in [3, 4]}_{int}
\end{equation}
Then, both '3c' and '2c' apply an e-IpDFT to the OOBI $\beta$ spectrum to estimate its parameters (line 30). This spectrum is the result of subtracting from $Y_{\beta_{H}}(k)$ the contribution of the fundamental component $(\hat{Y}_{0_{\beta}}^{q}(k))$ (line 29). $\hat{Y}_{0_{\beta}}^{q}(k)$ is calculated based on the $\beta$ parameters of the fundamental, ($\hat{A}_{0_{\beta}}^{q}$, $\hat{\varphi}_{0_{\beta}}^{q}$), both determined by taking into account the filter response (line 28). Accordingly, the undisturbed OOBI magnitude and phase ($\hat{A}_{i}^{q}$, $\hat{\varphi}_{i}^{q}$) are obtained by removing the filter effects from the estimates (lines 31-32). The spectral effects of both in-quadrature $\alpha$ and $\beta$ components on the positive and negative images of the OOBI tone are combined by means of the $\mathtt{DCSOGI\text{-}CG_{+-}}$ function called at line 33 in  Algorithm \ref{alg_2c3c_DCSOGI_IpDFT}, which must also consider the fundamental magnitude distortion correction. This function corresponds to \cite[Algorithm 3]{Veloso2023_SOGI_IpDFT_PT} and \cite[Algorithm 4]{Veloso2024_SOGI_IpDFT_SGSMA}. With the resulting gains $(\sigma_{\text{+}_{i}}^{q},\sigma_{\text{--}_{i}}^{q})$, the filtered spectrum of the interference $\hat{Y}_{i}^{q}(k)$ (lines 34-36) is calculated and used to improve the estimates in the next iteration of the loop.
As indicated in \cite{Veloso2024_SOGI_IpDFT_SGSMA}, the '2c' requires an additional correction if no OOBI is detected. This is because  2\textsuperscript{nd} harmonic tones with a magnitude below 10\% of the fundamental might not trigger the OOBI correction (see Section \ref{doc_sec_4} Fig. \ref{fig_OOBI_threshold_2c}(a)), and thus if left unattended would compromise the compliance with the $1\%$ harmonic disturbance requirements of class P. Despite the mitigation offered by the DCSOGI filter, the '3c' is also negatively affected by the presence of smaller 2\textsuperscript{nd} harmonic interferences under non-synchronous sampling conditions. However, this correction is indispensable for the '2c' to comply with \cite{PMU_Measurement_60255-118-1-2018}, as the adoption of a shorter observation window results in higher long-range leakage. Compared to \cite{Veloso2024_SOGI_IpDFT_SGSMA} a better resilience against 2\textsuperscript{nd} harmonic interferences is achieved in this work for both the '3c' and '2c' with the inclusion of a dedicated detection mechanism\footnote{In \cite{Veloso2024_SOGI_IpDFT_SGSMA} no specific mechanism was used and no further action was taken in the case of the '3c'. For the '2c' an additional 'always on' correction of a hypothetical 2\textsuperscript{nd} harmonic was imposed if no OOBI was detected.}. This mechanism also allows to prevent the false interpretation of other phenomena as 2\textsuperscript{nd} harmonic interferences which, in \cite{Veloso2024_SOGI_IpDFT_SGSMA}, limited the performance of the '2c' in the presence of amplitude modulations and steps. This detection is achieved by evaluating the ratio of $E_2$ (\ref{eq_E2_def}) to $E_i$, with $E_2$ referring to the energy content of those bins corresponding to a 2\textsuperscript{nd} harmonic within the residual spectrum $\hat{X}_{\varnothing_{i}}(k)$\footnote{To mitigate the interference of potential nearby tones bins [6-7] are considered for the '3c' and only bin 4 for the '2c' due to the higher spectral leakage associated with the shorter observation window.}. For the magnitudes considered in \cite{PMU_Measurement_60255-118-1-2018} an additional threshold $\lambda_2$ can be tuned so that the presence of a 2\textsuperscript{nd} harmonic interference can be detected (see Section \ref{doc_sec_4}). If so, that interference can be compensated analogously as an OOBI but with the e-IpDFT: (i) assuming the maximum bin corresponds to $k_i \text{ = } 2f_nT$ and (ii)  applied to the $\alpha$ spectrum instead of $\beta$ for the '2c' (lines 37-41).

Lastly, once the loop is exited i.e. $Q$ runs were completed or no interferences were detected, the estimation is concluded by removing the phase distortion introduced by the DCSOGI-QSG filter and shifting the estimated angle to the reporting instant $n_r$ (line 44). The ROCOF is calculated using consecutive frequency estimates via a first-order backward approximation of a first-order derivative \cite{Veloso2023_SOGI_IpDFT_PT}. It is worth noting that Algorithm \ref{alg_2c3c_DCSOGI_IpDFT} contains fundamental differences with respect to \cite[Algorithm 3]{Veloso2024_SOGI_IpDFT_SGSMA}. These are:
\begin{itemize}
    \item The use of the DCSOGI filter instead of the SOGI, as well as the adoption of the DC-Blocker, to provide the resulting SE with resilience against DC offsets.
    \item The explicit explanation of how the analysis window is repositioned to compensate for the group delay.
    \item The adoption of a new OOBI correction trigger for the '3c' based on the technique used in \cite{Veloso2023_TIM}.
    \item The inclusion of a dedicated 2\textsuperscript{nd} harmonic detection mechanism, for both the '3c' and '2c', to allow the identification and correction of low amplitude 2\textsuperscript{nd} harmonic tones. The mechanism also prevents the false interpretation of other phenomena as 2\textsuperscript{nd} harmonic interferences which previously affected and limited the performance of the '2c' in \cite{Veloso2024_SOGI_IpDFT_SGSMA}.
\end{itemize}
\begin{figure}[t]
\centering
\includegraphics[width=\linewidth]{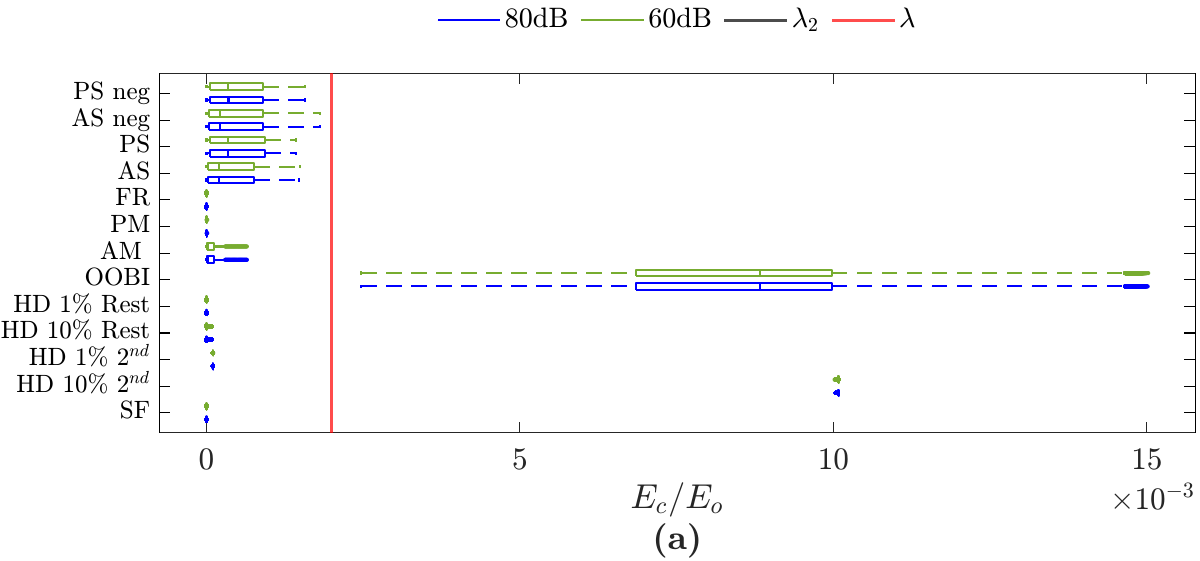}
\includegraphics[width=\linewidth]{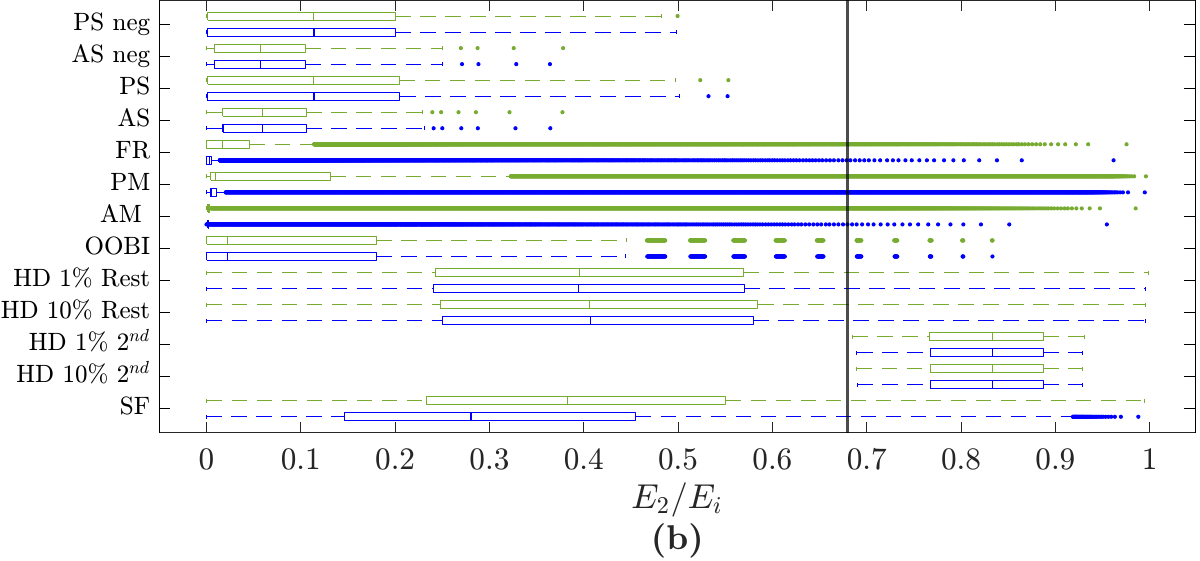}
\vspace{-5mm}
\caption{Boxplot representation for the '3c' algorithm under 80 and 60 dB AWGN of: (a) $E_i / E_o$ and (b) $E_2/E_i$ in all testing conditions in \cite{PMU_Measurement_60255-118-1-2018}.}
\label{fig_OOBI_threshold_3c}
\vspace{-5mm}
\end{figure}
\vspace{-2.5mm}\section{Interference Detection and $Q$ Tuning}\label{doc_sec_4}
This section presents the analysis conducted to adjust the 5 thresholds, $(\lambda,\lambda_2,\lambda_A,\lambda_\phi,\lambda_{int})$, involved in the OOBI and 2\textsuperscript{nd} harmonic detection across the '2c' and '3c' algorithms as well as the optimization of the maximum number of iterations, $Q$, required for each method. It is important to notice that compared to \cite{Veloso2024_SOGI_IpDFT_SGSMA,Veloso2023_SOGI_IpDFT_PT} all thresholds and $Q$ must be re-tuned due to the adoption of the new filters. All 5 thresholds have been derived by examining each test condition in \cite{PMU_Measurement_60255-118-1-2018} a total of 101 times, shifting the initial phase angles of the generated reference test signal between $[0;2\pi)$. In the case of the step tests, a total of 10093 initial angles are investigated instead to further analyze the effects of the relative position of the step within the waveform. Additionally an augmented reporting time of 2 ms ($F_r=500$ fps) has been considered\footnote{The use of the augmented $F_r$ is limited to the examination of a wider range of observation windows and angles within each test duration. The original $F_r$ value is kept as the 'de facto' value for all requirements in \cite{PMU_Measurement_60255-118-1-2018}.} together with two levels of additive white Gaussian noise (AWGN), namely 80 and 60 dB. Moreover, all tests to be performed at nominal frequency according to \cite{PMU_Measurement_60255-118-1-2018} are performed at 9 equidistant frequencies within the $[48 - 52]$ Hz interval to account for more realistic incoherent sampling\footnote{Incoherent sampling refers to the use of an observation window which does not contain an integer number of fundamental periods. Since the frequency is unknown, it represents the most likely operating condition.} conditions. Lastly more demanding testing conditions are also considered for the OOBI test, for which a comprehensive set of fundamental and interfering frequency values have been examined\footnote{A total of 7 fundamental frequencies $f_0$ are considered together with 69 interfering ones $f_i$. The former correspond to the nominal frequency as well as 3 pairs of equidistant ones within the $[47.5-49.5]$ Hz and $[50.5-52.5]$ Hz intervals. The latter are equally spaced using a 0.1 Hz and 1 Hz resolutions respectively in the $[10 - 11]\cup [24 - 25] \cup[75 - 76]$ Hz and $(11 -24)\cup(76 - 100]$ Hz intervals.}. For the adjustment of $Q$ the same OOBI testing conditions are used but limiting the examined $f_0$ to $f_0 \in [47.5,50,52.5]$ Hz. 

First, in the case of the '2c', the energy ratio $E_i/E_o$ (\ref{eq_Eo_def}) is examined by considering, as done in \cite{Veloso2024_SOGI_IpDFT_SGSMA}, only those bins up to and including the second harmonic. The results are shown in Fig. \ref{fig_OOBI_threshold_2c}(a), where now for better clarity the results of the 2\textsuperscript{nd} harmonic are explicitly separated from those of the remaining harmonics in the HD tests\footnote{The overlap between the HD 10\% 2\textsuperscript{nd} harmonic and the OOBI is not an issue as the former falls within the OOBI range.}. Variances $\text{Var}(y_{A}(n))/\hat{A}_0^2$ and $\text{Var}(y_{\phi}(n))$ are examined in Figs. \ref{fig_OOBI_threshold_2c}(b) and \ref{fig_OOBI_threshold_2c}(c) and the ratios of $E_{sub}$ and $E_{int}$ (\ref{eq_Eint_def}) with respect to $E_o$ depicted in Fig. \ref{fig_OOBI_threshold_2c}(d). In the case of the '3c' $\lambda$ is adjusted based on the ratio $E_c/E_o$ with the results shown in Fig. \ref{fig_OOBI_threshold_3c}(a). Finally, for both, the '2c' and '3c', the ratio $E_2/E_o$ is analyzed to identify potential 2\textsuperscript{nd} harmonic interferences when no OOBI is detected. As seen in Fig. \ref{fig_OOBI_threshold_2c}(e) and Fig. \ref{fig_OOBI_threshold_3c}(b) 2\textsuperscript{nd} harmonic tones exhibit, even for those with a low amplitude (HD $1\%$ 2\textsuperscript{nd}) a distinctively high $E_2/E_o$ ratio compared to the other tests. Moreover, a $\lambda_2$ can now be defined (for the magnitudes specified in \cite{PMU_Measurement_60255-118-1-2018}) to clearly discriminate between 2\textsuperscript{nd} harmonic interferences and AS. This allows to address the shortcomings of the 'always on' 2\textsuperscript{nd} harmonic correction which in \cite{Veloso2024_SOGI_IpDFT_SGSMA} negatively impacted the results of the AS tests.

\begin{figure}[!t]
\centering
\includegraphics[width=\linewidth]{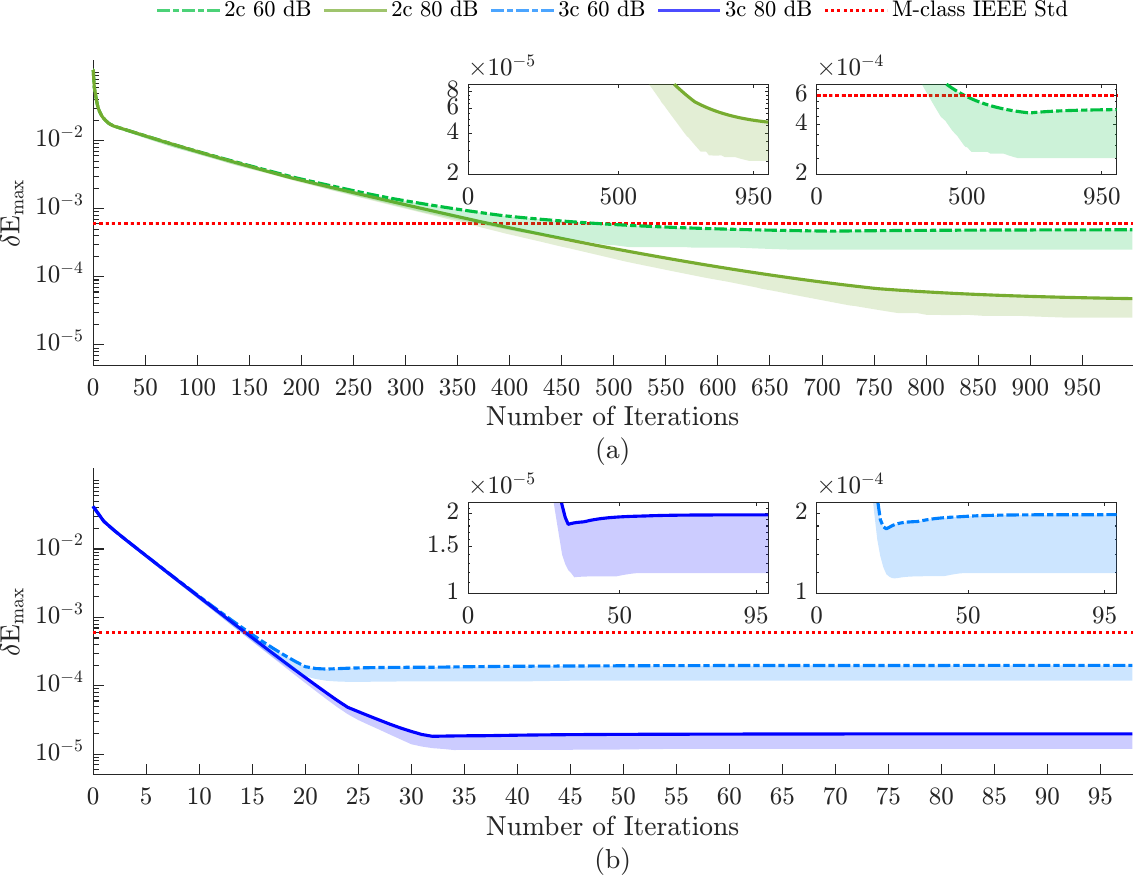}
\vspace{-5mm}
\caption{$\delta E_{\max}$ as a function of the total number of iterations for the '2c' (a) and '3c' (b) under 80 and 60 dB noise levels. The shaded areas represent the variability through the different considered cases, while the solid lines highlight the upper variability limits.}
\label{fig_Q_sel_OOBI}
\vspace{-5mm}
\end{figure}

As in \cite{Veloso2024_SOGI_IpDFT_SGSMA} $Q$ is again selected by means of the maximum overall error $\delta E_{\max}$, which is defined as:
\begin{equation}\label{eq_delta_ij_max}
\delta E_{\max} = \max_{f_{i}}{(\max_{f_{0}}{(\delta E_{f_{0}f_{i}})})}
\end{equation}where $\delta E_{f_{0}f_{i}}$ is the error in estimating the correction term $\delta$ \cite[eq.(3)]{Veloso2024_SOGI_IpDFT_SGSMA} for each frequency pair $f_0$-$f_i$. For simplicity, the same optimal number of internal iterations used in \cite{Veloso2024_SOGI_IpDFT_SGSMA} by the e-IpDFT are considered in the case of the '2c'. These were previously determined in \cite{Veloso2024_SOGI_IpDFT_SGSMA} with the conventional SOGI and correspond to a $P_i=3$. Thus, the analysis of $\delta E_{\max}$ can be restricted solely to a function of $Q$. Results for both methods are shown in Fig. \ref{fig_Q_sel_OOBI}, where shaded areas are used to represent the variability of the results through the considered cases for the '2c' (Fig. \ref{fig_Q_sel_OOBI}(a)) and the '3c' (Fig. \ref{fig_Q_sel_OOBI}(b)). A value of $Q=34$ is selected for the latter as it ensures the best performance under 80 dB noise. While for the '2c', to limit $Q$, a value of $Q = 711$ is selected for the same reason but under 60 dB noise, as with 80 dB $\delta E_{\max}$ is found to be monotonically decreasing within the considered range. Compared to \cite{Veloso2024_SOGI_IpDFT_SGSMA} the '2c' shows no violations of the M class limit for 60 dB. However, given the observed proximity, its full compliance under 60 dB cannot be completely ensured. Indeed, as shown in Section \ref{doc_sec_5}, some marginal violations might occur in such case. Nonetheless, as also previously pointed in \cite{Veloso2024_SOGI_IpDFT_SGSMA}, while \cite{PMU_Measurement_60255-118-1-2018} does not require the presence of noise to be considered, the adoption of both noise levels, of which 60 dB already represents a significantly challenging condition, allows us to explore and characterize the performance limitations of both methods.

\section{Performance Assessment} \label{doc_sec_5}
\begin{table}
\footnotesize
\begin{center}
\begin{threeparttable}
\caption{DCSOGI-IpDFT '3c' and '2c' parameters}
\label{tab_Param}
\begin{tabular}{c  c  c  c}
\toprule
Parameter & Variable & 3 cycle '3c' & 2 cycle '2c'\\
\midrule
Nominal System Frequency & $f_{n}$ & 50 Hz & 50 Hz\\
Window Type & -& Hann & Hann\\
Window Length & $T$ & 60 ms $(\frac{3}{f_{n}})$ & 40 ms $(\frac{2}{f_{n}})$\\
Sampling Rate & $F_{s}$ & 50 kHz & 50 kHz \\
PMU Reporting Rate & $F_{r}$ & 50 fps & 50 fps\\
DFT bins & $K$ & 8 & 6\\
Self-Inter. comp.   &$P_{i}$ & 2 & 3\\
Max Number of Iterations & $Q$ & 34 & 711\tnote{a}\\
IpDFT Interpolation Points & -& 3 & 3\\
OOBI Detection Threshold& $\lambda$ & $2 \cdot 10^{-3}$ & $5.5 \cdot 10^{-4}$\\
Envelope Threshold& $\lambda_{A}$ & - & $1.2 \cdot 10^{-3}$\\
Angle Threshold& $\lambda_{\phi}$ & - & $3 \cdot 10^{-8}$\\
Interharmonic Threshold& $\lambda_{int}$ & - & $7.5 \cdot 10^{-4}$\\
2\textsuperscript{nd} Harm. Energy Con. Th.& $\lambda_{2}$ & $6.8 \cdot 10^{-1}$ & $7 \cdot 10^{-1}$\\
Settling Time & $t_{s}$& 20 ms & 20 ms\\
Filter Centre Frequency & $\omega_{c}$& $2\pi f_{n}$ & $2\pi f_{n}$\\
SOGI-QSG Gain &$k_{s}$ & $9.2/(t_{s}\omega_{c})$ & $9.2/(t_{s}\omega_{c})$\\
DCSOGI-DC Gain &$k_{\upsilon}$ & $0.2104$ & $0.2104$\\
DC Blocker pole &$p$ & $0.999$ & $0.999$\\
\bottomrule
\end{tabular}
\begin{tablenotes}
\item[a] A smaller value of $Q=18$ is used for the '2c' instead when the 2\textsuperscript{nd} harmonic mechanism is triggered as it was found to be sufficient to compensate such interferences.
\end{tablenotes}
\end{threeparttable}
\end{center}
\vspace{-5mm}
\end{table}

Both the '2c' and '3c' algorithms are here evaluated against the P and M accuracy limits defined in \cite{PMU_Measurement_60255-118-1-2018}. A MATLAB simulated testbed is used and a characterization of the two techniques is carried out in terms of total vector error (TVE), frequency error (FE), and ROCOF error (RFE), as well as response times ($R_{t}$), delay times ($D_{t}$) and maximum overshoot values (OS) for the step tests. To showcase the methods' resilience against DC offsets, all tests in \cite{PMU_Measurement_60255-118-1-2018} have been superposed with a static DC component whose magnitude equals 10\% of that of the fundamental tone. Both '2c' and '3c' are tested on the basis of the parametrization given in Table \ref{tab_Param}. Test reference signals considering different initial phase angles\footnote{Note that a different numbers of angles are selected compared to those used for the tuning of the different thresholds in Section \ref{doc_sec_4}. This is done to avoid testing the methods across the same observation windows and ensure a fair assessment.} and affected by two levels of AWGN (60 and 80 dB) are generated and evaluated. The worst-case results among all the evaluated scenarios are then plotted in Figs. \ref{fig_SS_All} - \ref{fig_DS_All} for the static and dynamic tests together with the Std. accuracy limits \cite{PMU_Measurement_60255-118-1-2018}. Moreover, the maximum values of the step tests are presented in Table \ref{tab_Steps} together with a detailed evaluation of their OS and $D_t$ shown in Fig. \ref{fig_Steps_detail}. Lastly a final set of tests, beyond \cite{PMU_Measurement_60255-118-1-2018}, employing signals corrupted by a custom harmonic profile based on the maximum individual harmonic limits for distribution networks defined in \cite{Supply_Voltage_Quality_Std_50160_2022} have been considered. 

\subsection{Static Tests}
\begin{figure}[!t]
\centering
\includegraphics[width=\linewidth]{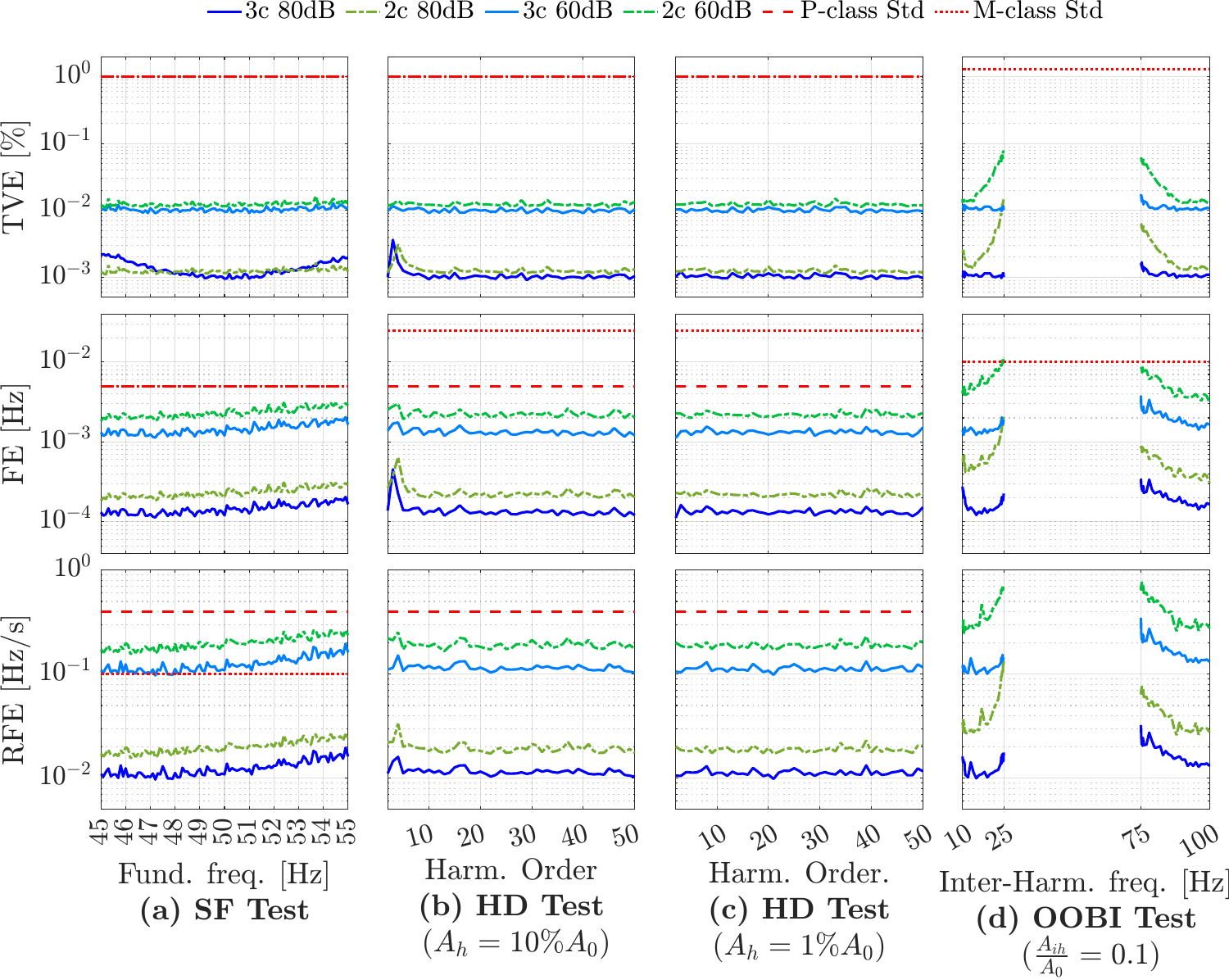}
\vspace{-7mm}
\caption{Static tests: (a) Signal frequency range (SF), (b) $10\%$ Harmonic distortion (HD) ($A_{h} = 10\% A_{0}$), (c) $1\%$ Harmonic distortion (HD) ($A_{h} = 1\% A_{0}$) and (d) OOBI ($A_{ih} = 10\% A_{0}$) \cite{PMU_Measurement_60255-118-1-2018}.}
\label{fig_SS_All}
\vspace{-2.5mm}
\end{figure}

All cases consider 151 different 1 s test signals generated by shifting their initial absolute phase angle by $2\pi/151$. Additionally the same augmented $F_r = 500$ fps used in Section \ref{doc_sec_4} is adopted for all tests. Again, this is limited in scope simply to examine a greater number of windows within the test duration, with the original $F_r = 50$ fps kept for all requirements in \cite{PMU_Measurement_60255-118-1-2018} as well as for the ROCOF calculation. Furthermore, the same fundamental $f_0$ and interfering frequencies $f_i$ used in Section \ref{doc_sec_4} for the selection of $Q$ are considered for the OOBI tests while the HD tests are conducted with a $f_0=49$ Hz to impose incoherent sampling conditions as well as an increased spectral proximity between the harmonic tones. Maximum errors among all windows are summarized in Fig. \ref{fig_SS_All}.  As expected, the '2c' implementation cannot match the overall accuracy of the '3c' variant regardless of the noise level. The sole exceptions being for the 80 dB noise case: (i) the signal frequency range test (SF)  when significant deviations from $f_n$ exists in terms of TVE and (ii) the HD tests in terms of TVE and FE when a 3\textsuperscript{rd} harmonic is present. Both methods comply with the requirements of P and M classes for 60 and 80 dB for the HD tests (Figs. \ref{fig_SS_All}(b) - \ref{fig_SS_All}(c)), with the novel 2\textsuperscript{nd} harmonic detection mechanism allowing them to meet the Std. requirements even under the currently imposed severe testing conditions. In the case of the OOBI test (Fig. \ref{fig_SS_All}(d)) the '2c' marginally exceeds the M class FE limit under 60 dB. However note that this represents a very demanding noise level under which, as shown in Fig. \ref{fig_SS_All}(a), neither method can meet the RFE M class limit for the SF test.
\subsection{Dynamic Tests}
\begin{figure}[!t]
\centering
\includegraphics[width=\linewidth]{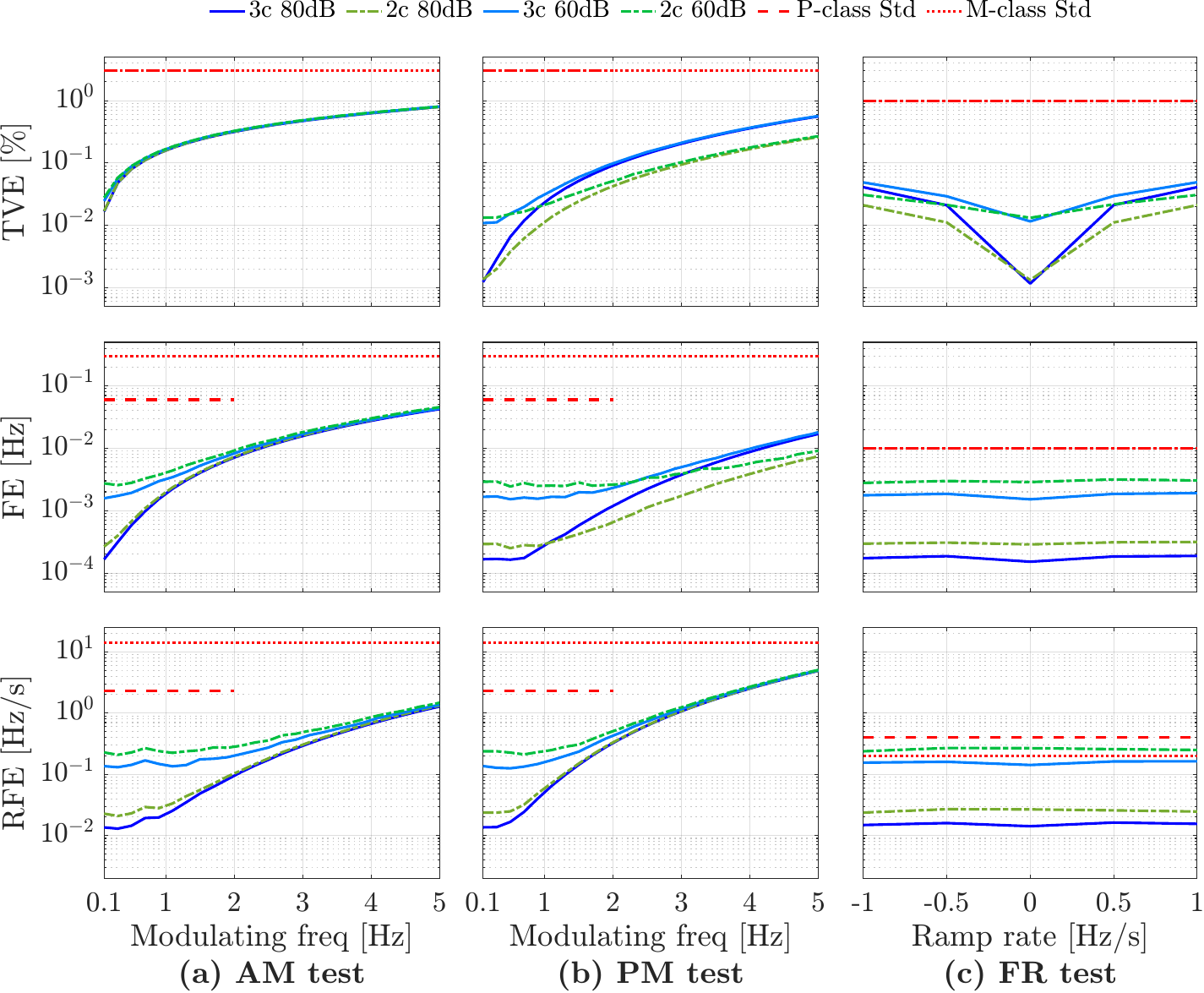}
\vspace{-7mm}
\caption{Dynamic tests: (a) Amplitude modulation (AM) (depth 10\%); (b) Phase modulation (PM) (depth $\pi/18$ rad); and (c) Frequency ramp (FR) \cite{PMU_Measurement_60255-118-1-2018}.}
\label{fig_DS_All}
\vspace{-5mm}
\end{figure}

Again 151 tests are conducted shifting the initial absolute phase angle of the test signal by $2\pi/151$ with the same augmented $F_r = 500$ fps used in Section \ref{doc_sec_4}. The test duration is $\max(\lceil2/f_m\rceil,5)$ s for all bandwidth tests i.e. amplitude (AM) and phase modulation (PM) and $12/|R_f|$ s for the ramp tests (FR), where $f_m$ and $R_f$ denote the modulating frequency and the ramp rate. As in the static tests, for the AM and PM tests (Figs. \ref{fig_DS_All}(a) - \ref{fig_DS_All}(b)), the '3c' outperforms the '2c' variant regardless of the noise level in terms of FE and RFE. This can also be observed for the FR (Fig. \ref{fig_DS_All}(c)) regardless of the ramp rate. In fact, for the latter, the '2c' exceeds again the M-class limit under 60 dB. However, with increasing $f_m$, the '2c' implementation can either deliver more accurate estimates (see the TVE and FE for the PM test) or match the performance of the '3c' (see all metrics for the AM test and RFE for the PM test). Similarly, in the FR test and in terms of TVE, the '2c' becomes more accurate with increasing ramp rates. Overall, $f_m$ becomes the main source of error over noise for the higher $f_m$ values for both cases. Likewise, for the FR test, while the FE and RFE remain largely unaffected by $|R_f|$, its influence on the TVE is apparent, with increasing $|R_f|$ resulting in higher TVEs. Finally compared to \cite{Veloso2024_SOGI_IpDFT_SGSMA}, a higher accuracy is also achieved for the '2c' during the AM tests thanks to the new 2\textsuperscript{nd} harmonic detection technique.
\subsection{Step Tests}
\begin{figure}[!t]
\centering
\includegraphics[width=\linewidth]{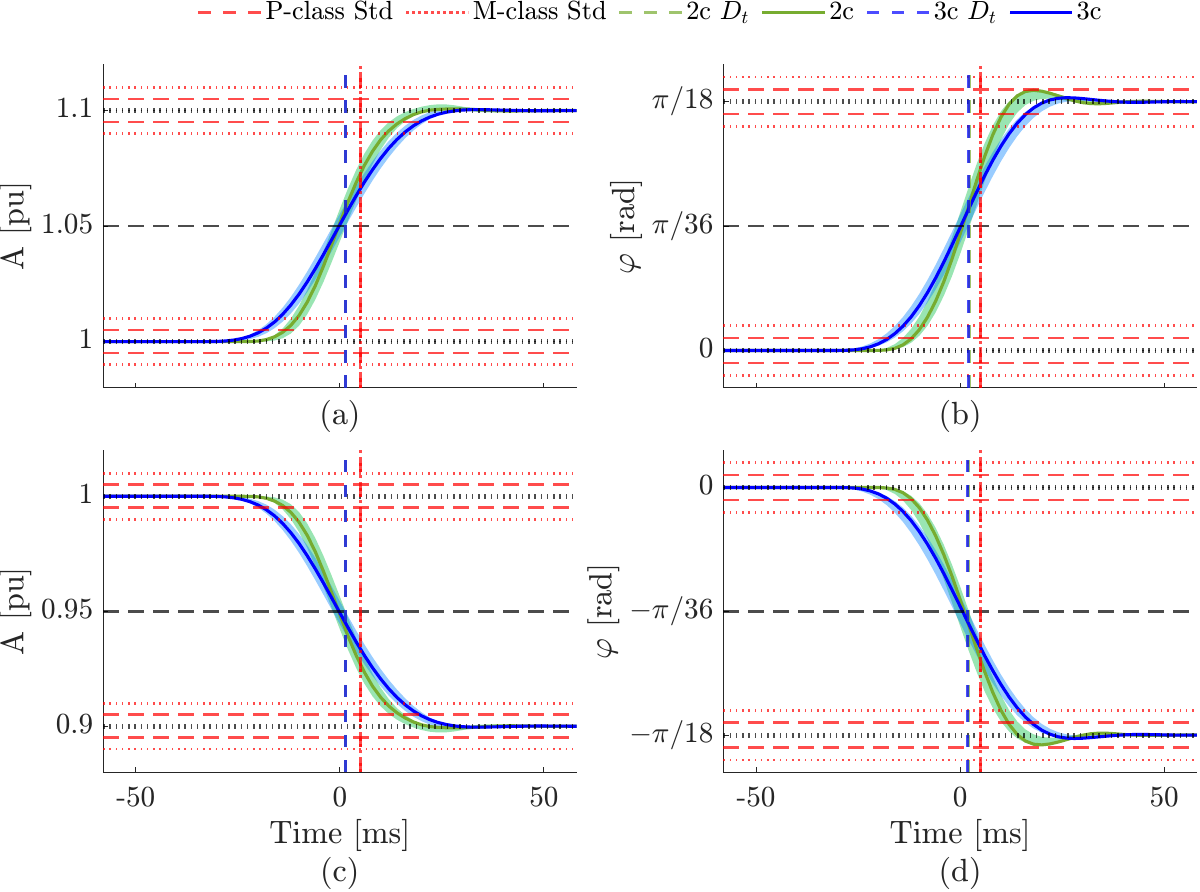}
\vspace{-5mm}
\caption{Detailed OS and $D_t$ evaluation: (a) Positive AS tests (+10\%), (b) Positive PS tests ($+\pi/18$), (c) Negative AS tests (-10\%) and (d) Negative PS tests ($-\pi/18$). The shaded areas represent the envelopes of the trajectories of all the cases considered and the solid lines are used to illustrate one of such trajectories (blue for the '3c' and green for the '2c'). Auxiliary black lines represent the state boundaries (dotted lines) and the mid-point value between the states (dashed line). The plots are centered at the instant the step occurs.}
\label{fig_Steps_detail}
\vspace{-5mm}
\end{figure}

A test duration of 1.5 s is considered for all step tests with the step occurrence fixed at 1 s. A total of 10001 different reference signals are considered by shifting their initial absolute phase angle by $2\pi/10001$. This allows to modify the relative position of the step so that a comprehensive assessment can be done. The maximum obtained values among the positive and negative amplitude (AS) and phase (PS) steps are summarized in Table \ref{tab_Steps} for both 60 and 80 dB cases. A shortcoming of the 'always on' 2\textsuperscript{nd} harmonic correction used by the '2c' in \cite{Veloso2024_SOGI_IpDFT_SGSMA} was the impact caused across the step tests. Indeed, during the transient, the spread of the spectral energy caused by the step was misinterpreted as a 2\textsuperscript{nd} harmonic tone which manifested as a higher maximum delay time on the amplitude steps. To assess the efficiency of the novel 2\textsuperscript{nd} harmonic detection, as also done in \cite{Veloso2024_SOGI_IpDFT_SGSMA}, the envelopes of all trajectories of the estimated fundamental amplitudes and phase differences for all the cases considered are presented in Fig. \ref{fig_Steps_detail} by means of shaded areas. These allow to visually examine whether a potential false correction might have been performed and its resulting impact. For better clarity only the 80 dB cases are shown in Fig. \ref{fig_Steps_detail} (similar results are obtained under 60 dB noise).

Table \ref{tab_Steps} shows how the use of a shorter observation window results in shorter $R_t$ compared to the '3c'. Note that the RFE $R_t$ under 60 dB is not reported. This is because as shown in Fig. \ref{fig_SS_All}(a), under 60 dB the RFE achieved by both methods already exceed the 0.1Hz/s M-class limit. The adoption of the DCSOGI also results in higher maximum overshoots compared to \cite{Veloso2024_SOGI_IpDFT_SGSMA}, once again with the '2c' presenting higher values than the '3c'. However, these are within standard requirements. Finally, in terms of delay time, the 2\textsuperscript{nd} harmonic detection mechanism allows to correct the increased $D_t$ values observed in \cite{Veloso2024_SOGI_IpDFT_SGSMA} during the AS tests. As shown in Fig. \ref{fig_Steps_detail} a smooth transition is obtained across all step cases and the resulting maximum $D_t$ values are well within the standard limits.
\begin{table}
\footnotesize
\begin{center}
\begin{threeparttable}
\caption{Maximum $R_t$, $D_t$ and OS in Step Tests and Limits Allowed by \cite{PMU_Measurement_60255-118-1-2018}}
\label{tab_Steps}
\begin{tabularx}{0.49\textwidth}{m{0.1cm} c >{\centering\arraybackslash}X >{\centering\arraybackslash}X | c >{\centering\arraybackslash}X >{\centering\arraybackslash}X}
\toprule
   \textit{SNR} &  Std&  '3c'&  '2c'&  Std&  '3c'& '2c'\\
    \textit{dB}&  P$\setminus$M&  \textit{60}$\setminus$\textit{80}&  \textit{60}$\setminus$\textit{80}&  P$\setminus$M&  \textit{60}$\setminus$\textit{80}& \textit{60}$\setminus$\textit{80}\\
     \midrule
 & \multicolumn{3}{c}{TVE $R_t$ [ms]}& \multicolumn{3}{c}{FE $R_t$ [ms]}\\
     \midrule
     AS&  40$\setminus$140&  33.8$\setminus$33.8&  26.2$\setminus$26.2&  90$\setminus$280&  74.6$\setminus$68.6& 70.6$\setminus$66.7\\
    PS&  40$\setminus$140&  37.0$\setminus$37.0&  31.9$\setminus$31.9&  90$\setminus$280&  77.0$\setminus$76.2& 73.4$\setminus$71.8\\
    \midrule
     & \multicolumn{3}{c}{RFE $R_t$ [ms]\tnote{a}}& \multicolumn{3}{c}{$D_t$ [ms]}\\
    \midrule
     AS&  120$\setminus$280&  -$\setminus$100.5&  -$\setminus$95.8&  5$\setminus$5&  1.4$\setminus$1.4& 1.5$\setminus$1.5\\
    PS&  120$\setminus$280&  -$\setminus$105.0&  -$\setminus$97.7&  5$\setminus$5&  2.0$\setminus$2.0& 2.3$\setminus$2.3\\
    \midrule
     & \multicolumn{3}{c}{Max OS AS [\%]}& \multicolumn{3}{c}{Max OS PS [\%]}\\
    \midrule
     &  5$\setminus$10&  1.4$\setminus$1.4&  2.8$\setminus$2.7&  5$\setminus$10&  1.6$\setminus$1.6&  4.6$\setminus$4.6\\

\bottomrule
\end{tabularx}
\begin{tablenotes}
\item[a] The RFE $R_t$ have been calculated considering all crossings with the M class limit within a 152 ms window around the step. This is done to exclude spurious RFE values that could marginally exceed said limit solely due to noise. No RFE values are reported under 60 dB noise as both methods were already shown to exceed the 0.1 Hz/s M class limit in Fig. \ref{fig_SS_All}(a).
\end{tablenotes}
\end{threeparttable}
\end{center}
\vspace{-5mm}
\end{table}

\subsection{Multiple Interferences}
\begin{figure}[!t]
\centering
\includegraphics[width=\linewidth]{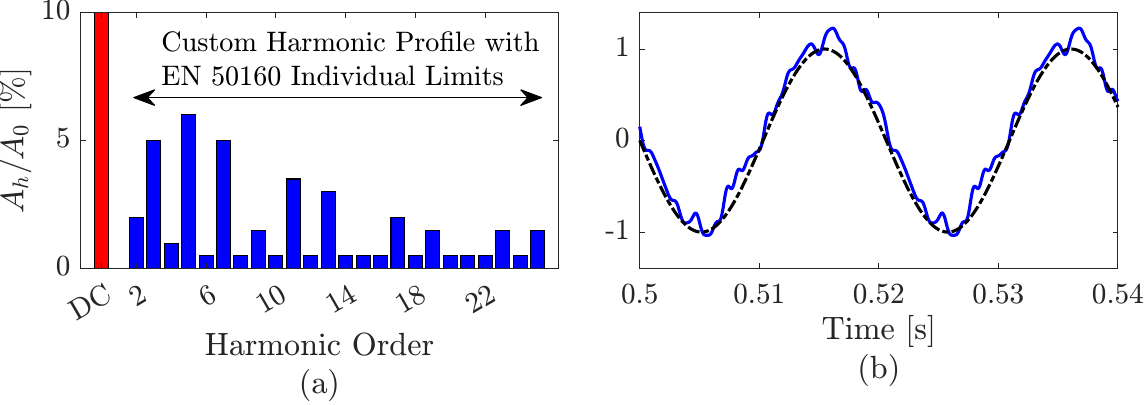}
\includegraphics[width=\linewidth]{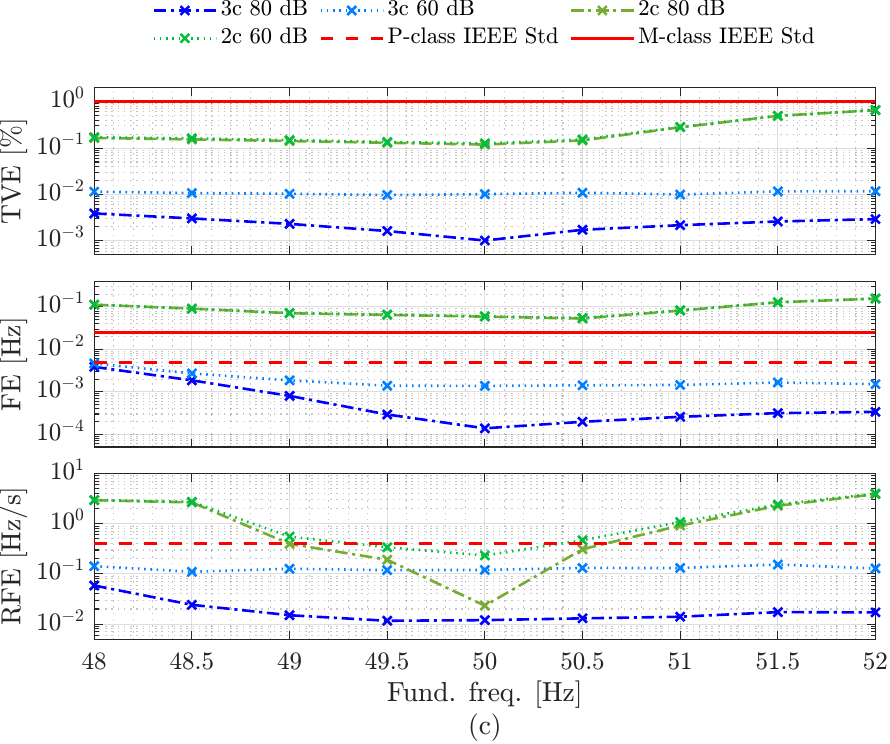}
\vspace{-5mm}
\caption{Total harmonic content considered for the multiple interference tests based on the maximum individual values for each harmonic tone according to the EN 50160 Std. \cite{Supply_Voltage_Quality_Std_50160_2022} (a); resulting sample test signal (blue line) together with the underlying fundamental tone (black line) (b), and worst case results across all considered cases by fundamental frequency (c). P and M class HD accuracy limits \cite{PMU_Measurement_60255-118-1-2018} simply provided as reference.}
\label{fig_Mult_Int}
\vspace{-5mm}
\end{figure}

To evaluate the methods' resilience against multiple simultaneous interferences (beyond the scope of \cite{PMU_Measurement_60255-118-1-2018}), a signal corrupted by a custom harmonic profile derived from the maximum values for each individual harmonic indicated in the EN 50160 Std. \cite{Supply_Voltage_Quality_Std_50160_2022} for power distribution networks (see Fig. \ref{fig_Mult_Int}(a)) has been considered\footnote{Note that the considered profile is beyond the requirements of \cite{Supply_Voltage_Quality_Std_50160_2022} where it is indicated that the total harmonic distortion (THD) must be $\leq 8\%$.}. A similar profile, but limited from the 2\textsuperscript{nd} to the 7\textsuperscript{th} harmonic, has also been previously used in \cite{Macii2021_IpDFT_Tuned_Compar,Macii2024_SingleCycle_P_class}. Moreover, for the current analysis, a 10\% static DC offset has again been superposed on top of the harmonic profile (see the red bar in Fig. \ref{fig_Mult_Int}(a)). 

The same test cases and testing conditions used for the static tests are applied. Additionally for each test the initial phase angles of all harmonic tones are randomly assigned between $0$ and $2\pi$ following an uniform distribution. All tests are then repeated for a total of 9 equidistant fundamental frequencies within the $[48-52]$ Hz range. A sample test signal is provided in Fig. \ref{fig_Mult_Int}(b) (blue line) for a 60 dB analysis case as reference. The worst case results for all considered cases are presented in Fig. \ref{fig_Mult_Int}(c) by fundamental frequency. The static HD accuracy limits from \cite{PMU_Measurement_60255-118-1-2018} are additionally provided for reference.

Results show how the '3c' even under such distorted conditions can provide reliable and accurate estimates across the considered fundamental frequency range regardless of the noise level. Furthermore, it is able to meet the HD requirements in \cite{PMU_Measurement_60255-118-1-2018} which are meant for a single noiseless harmonic disturbance under nominal system frequency conditions.

As expected, in the case of the '2c', the harmonic mitigation offered by the DCSOGI and the side lobe decay of the Hanning window are not enough to compensate the effects of the mutual spectral interference between consecutive harmonic tones. Given the coarser frequency resolution due to the shorter window length, the width of the main lobe of the Hanning window causes significantly overlap between adjacent tones. This, in turn, compromises the necessary removal of the 2\textsuperscript{nd} harmonic, which is essential for an efficient estimation of the fundamental component for the '2c'. However, it is important to note that the method is still capable of providing phasor estimates with a TVE below 1\%. 

\section{Conclusions}\label{doc_sec_6}
This paper presented an evolution of the SOGI-IpDFT SE, the DCSOGI-IpDFT, capable of maintaining simultaneous compliance across all Std. tests with classes P and M even in the presence of DC offsets. Once again, two variants using two ('2c') and three ('3c') nominal fundamental period windows were formulated,  relying on the same three-point interpolation technique, Hanning window, and e-IpDFT algorithm as the SOGI-IpDFT. The method further incorporated a dedicated mechanism for the detection and correction of low amplitude 2\textsuperscript{nd} harmonic tones to ensure Std. requirements can be met in the presence of such disturbances even under off-nominal frequency conditions. This mechanism also allowed to address some of the shortcomings of the SOGI-IpDFT '2c' variant by preventing the misinterpretation of other phenomena as 2\textsuperscript{nd} harmonic interferences which previously limited its performance. 

Overall the '2c' provides less accurate estimates than the '3c' across all static tests and through most of the dynamic ones, with the exception of the PM tests under high modulating frequencies and the FR tests in terms of TVE. However, the '2c' can also achieve shorter response times during the steps in exchange for larger overshoots. Both methods comply with all standard tests, including the OOBI, even in the presence of a 10\% DC offset, with only violations of a few M class limits under the challenging conditions posed by the 60 dB noise level. Once again, to comply with the OOBI test, the '2c' requires the use of a larger number of user-defined thresholds and of iterations compared to the '3c'. As it was the case for the SOGI-IpDFT, a generic OOBI with an amplitude lower than 10\% might not necessarily be detected by neither variant and thus in such case the same level of accuracy cannot be ensured. Lastly it is important to highlight the advantage the '3c' has when multiple harmonic interferences are present due to the longer observation window. The method showed remarkable performance in the conducted tests being even able to meet the HD standard requirements, making it the preferred choice if signals with high total harmonic distortion are expected.\vspace{-5mm}
\printbibliography
\end{document}